\documentclass[a4paper,11pt]{article}
\usepackage{jcappub} 
\usepackage{packages}

\arxivnumber{2603.06792} 
\title{Spectral sirens cosmology from binary black holes populations with sharper mass features}

\author[a, b]{Tom Bertheas\footnote{Corresponding author},}
\author[a]{Vasco Gennari,}
\author[b]{Danièle A.~Steer}
\author[a]{and Nicola Tamanini,}
\affiliation[a]{L2IT, Université de Toulouse, CNRS/IN2P3, Toulouse, France}
\affiliation[b]{{Laboratoire de Physique de l'ENS, Universit\'e Paris Cit\'e, Ecole Normale Sup\'erieure, Universit\'e PSL, Sorbonne Universit\'e, CNRS, 75005 Paris }}

\emailAdd{tom.bertheas@l2it.in2p3.fr, vasco.gennari@l2it.in2p3.fr, daniele.steer@phys.ens.fr, nicola.tamanini@l2it.in2p3.fr}

\abstract{
Spectral-sirens inference enables the extraction of cosmological parameters from gravitational-wave data alone, without electromagnetic counterparts or galaxy catalogs.
We introduce new parametric mass functions for the binary black hole population built as linear combination of truncated power-laws that capture significant structure across the mass spectrum.
On analysing the latest gravitational-wave transient catalog, GWTC-4.0, we show that power-laws-only population models constrain the Hubble constant to $H_0 = 53.3^{+14.0}_{-10.8} ~\rm km \,s^{-1} \,Mpc^{-1}$ at $68\%$ confidence level.
After probing the robustness of the results with respect to several modelling assumptions, we further test alternative cosmological models, establishing competitive constraints on modified gravitational-wave propagation, while bounds on the dark energy equation-of-state parameters remain uninformative.
Projecting to the future O5 observing run with larger datasets at higher redshifts, we forecast substantial improvements in $H_0$ and modified propagation parameters.
Our results highlight the strong interplay between the black hole mass distribution and inferred cosmology.}

\begin{document}
\maketitle
\flushbottom


\newpage

\section{Introduction}
\label{sec:intro}

Since their first direct detection in 2015 \cite{LIGOScientific:2016aoc}, \acp{GW} have become a new promising avenue to probe the universe, independent but complementary to \ac{EM} signals. 
In the last decade, interferometers of the \ac{LVK} collaboration~\cite{LIGOScientific:2014pky, VIRGO:2014yos, KAGRA:2020tym} have recorded hundreds of signals sourced by \acp{CBC}, including mostly \acp{BBH} as well as a few \acp{BNS} and composite binary systems (NSBH) \cite{LIGOScientific:2018mvr, LIGOScientific:2021usb, KAGRA:2021vkt, LIGOScientific:2025slb}. 
The plethora of sources thus observed enabled statistical studies of their astrophysical distribution in terms of the binaries' individual properties \cite{LIGOScientific:2020kqk, KAGRA:2021duu, LIGOScientific:2025pvj} and tests of \ac{GR} in the strong field regime \cite{LIGOScientific:2019fpa, LIGOScientific:2020tif, LIGOScientific:2021sio, LIGOScientific:2016lio, LIGOScientific:2018dkp, LIGOScientific:2025rid, LIGOScientific:2025wao}.

Most importantly for the present work, \ac{GW} observation can also provide valuable information on cosmological parameters \cite{LIGOScientific:2017adf, LIGOScientific:2019zcs, DES:2019ccw, LIGOScientific:2021aug, LIGOScientific:2025jau}, an essential asset in a time of growing tensions in cosmology that challenge the standard model of cosmology (namely \LCDM) \cite{DiValentino:2021izs, Perivolaropoulos:2021jda}. 
In particular, \acp{GW} carry direct information on the luminosity distance of the source, independent of any distance calibration, hence are usually referred to as \textit{standard sirens}~\cite{Schutz:1986gp, Markovic:1993cr, Holz:2005df}, mirroring \textit{standard candles} in astronomy. 
However, a single \ac{GW} observation from a \ac{CBC} is not sufficient to extract cosmological information because of the perfect mass-redshift degeneracy.

As a consequence, \ac{GW} cosmology currently relies mostly on methods combining information from multiple detected events, which require a careful modelling of the astrophysical distribution of \acp{CBC} in terms of their individual properties  such as masses, distance, spins, etc,~(see~\cite{Palmese:2025zku, Pierra:2025fgr} for detailed reviews). 
Taking advantage of the presence of prominent features in the \acp{CBC} mass spectrum to statistically break the mass-redshift degeneracy, the so-called \textit{spectral sirens} method allows one to infer cosmological parameters together with population ones~\cite{Chernoff:1993th, Finn:1995ah, Taylor:2011fs, Taylor:2012db, Farr:2019twy, Ezquiaga:2022zkx, Mastrogiovanni:2021wsd}, under some assumptions on the intrinsic  evolution of astrophysical distribution~\cite{Mukherjee:2021rtw, Pierra:2023deu, Agarwal:2024hld, Gennari:2025nho}. 
Informing redshift priors from EM data in the form of galaxy catalogs can further improve the resulting cosmological constraints, a method often called \textit{dark sirens}~\cite{Schutz:1986gp, Holz:2005df, Gair:2022zsa, Gray:2023wgj}. 
The ideal case occurs when a \ac{GW} event is accompanied by an \ac{EM} counterpart (a \textit{bright siren}), enabling unique identification of the host galaxy and precise redshift measurements via spectroscopic surveys~\cite{LIGOScientific:2017adf, LIGOScientific:2018hze, Palmese:2023beh}. 
However, the lack of such exceptional events, as well as the low completeness of current all-sky galaxy catalogs~\cite{LIGOScientific:2025jau}, make spectral sirens of paramount importance for \ac{GW} cosmology. 
The present paper aims at presenting state-of-the-art spectral sirens cosmological constraints.

The spectral siren method hinges on a suitable model of the mass distribution of GW sources, able to leverage all mass features without introducing systematic bias.
Multiple independent analyses of most recent data support robust and model-independent evidence for the presence of pronounced overdensities in the \ac{BBH} primary mass spectrum at $\sim 10 \,M_\odot$ and $\sim 35 \,M_\odot$~\cite{LIGOScientific:2025pvj, LIGOScientific:2025jau, Guttman:2025jkv, Sridhar:2025kvi, Tiwari:2025oah, Mali:2024wpq}, which are the main \ac{BBH} mass features exploited for \ac{LVK} spectral sirens results~\cite{LIGOScientific:2025jau}. Moreover, several studies describing the mass spectrum both in a parametric and non-parametric fashion have reported evidence of additional features, both in the low ($\sim 10 - 20 \,M_\odot$)~\cite{Toubiana:2023egi, Gennari:2025nho, Tiwari:2025lit, Tiwari:2025oah, Willcox:2025poh, Bertheas:2025mzd} and high ($\gtrsim 50 \,M_\odot$)~\cite{MaganaHernandez:2024qkz, Pierra:2026ffj, Sridhar:2025kvi} mass regions, and related improved constraints on $H_0$.
Here, we propose new mass models that leverage this structure in a fully parametric fashion.

Challenges faced by the concordance \LCDM model motivate the current paper in two ways. 
First, \acp{GW} represent a promising cosmological messenger that is independent of other probes: it is therefore essential to develop and improve current methods to get new insights on the universe and its properties.
To that end, we propose the use of new population models to constrain cosmological parameters within \LCDM. 
Secondly, we introduce more general cosmological models and constrain potential deviations from \LCDM. 
Within \ac{GR}, the current knowledge on the energy content of the universe is questioned, with recent findings from the \ac{DESI} supporting evidence for a time-evolving \ac{DE} component~\cite{DESICollaboration2025ddr}.
If prospects have been proposed for next generation \ac{GW} detectors~\cite{Du:2018tia, Afroz:2024lou, Petiteau:2011we}, we investigate our ability to constrain such models using current data.
\ac{GW} also naturally carry information on potential deviation from \ac{GR}; multiple alternative theories of gravity thus predict modifications of \ac{GW} propagation on cosmological scales~\cite{Clifton:2011jh, Joyce:2014kja}, which can also be constrained by \ac{GW} data~\cite{LIGOScientific:2025jau, Leyde:2022orh, Chen:2023wpj, Tagliazucchi:2025ofb} 

The aims of the paper are twofold. 
The first is to re-analyse data from the latest \ac{GWTC} with more flexible parametric mass model than those considered in \cite{LIGOScientific:2025jau}, which we do in Sec.~\ref{sec:res:gwtc4}. 
We show that in the context of \ac{GR} our models -- describing the primary mass distribution as a linear combination of tapered power laws -- capture different population phenomenology compared to fiducial models found in the literature, and lead to smaller relative uncertainty on $H_0$. 
We additionally extend the analysis to consider theories of gravity beyond \LCDM, reporting no significant measurements of dynamical \ac{DE} parameters, and providing the best constraints to date on modified gravity parameters using the \ac{BBH} population alone. 
The second aim is to provide forecasts for the future O5 observing run, as presented in Sec.~\ref{sec:res:O5}. 
To that purpose, we create a mock dataset using the best fit mass model from \ac{GWTC}, and produce a most realistic simulation at O5 sensitivity, including full event parameter estimation in detector noise.
We describe our framework in Sec.~\ref{sec:methods}.

\section{Methods}
\label{sec:methods}
\subsection{Hierarchical inference framework}

Our goal is to infer a set of \textit{population-level} parameters $\bm{\Lambda}$ --- e.g. the minimum and maximum primary mass of the \acp{BBH} population, mass scales of overdensities, cosmological parameters, etc. --- from an ensemble of $N_{\rm obs}$ \ac{GW} observations associated with the data $(\bm{x}_1, \dots, \bm{x}_{N_{\rm obs}})$.
We use a \textit{hierarchical Bayesian inference} framework to calculate the posterior distribution $p(\bm{\Lambda} | \{ \bm{x} \})$ taking into account selection effects due to the finite sensitivity of our detectors \cite{Mandel:2018mve, Gaebel:2018poe, Vitale:2020aaz}:
\begin{equation}
\label{eq:hierarchical_likelihood}
    p(\bm{\Lambda} | \{ \bm{x} \}) \propto \pi(\bm{\Lambda}) \prod_{i=1}^{N_{\rm obs}} \frac{\int \mathcal{L}_{\rm ev}(\bm{x}_i | \bm{\theta}_d) \frac{\dd \mathcal{N}}{\dd \bm{\theta}_d \dd t_d}(\bm{\theta}_d | \bm{\Lambda}) \dd \bm{\theta}_d}{\int p_{\rm det}(\bm{\theta}_d) \frac{\dd \mathcal{N}}{\dd \bm{\theta}_d \dd t_d} (\bm{\theta}_d | \bm{\Lambda}) \dd \bm{\theta}_d}.
\end{equation}
Here $\pi(\bm{\Lambda})$ is a prior distribution on $\bm{\Lambda}$, $\mathcal{L}_{\rm ev}(\bm{x}_i | \bm{\theta}_d)$ is the single event likelihood of a \ac{BBH} with detector-frame\footnote{In the following, subscripts $d$ and $s$ refer to quantities evaluated in the detector- and source-frame respectively.} parameters $\bm{\theta}_d$ --- e.g. masses, distance, spins, sky localisation, etc. --- producing data $\bm{x}_i$ in the detectors, $\frac{\dd \mathcal{N}}{\dd \bm{\theta}_d \dd t_d}(\bm{\theta}_d | \bm{\Lambda})$ models the differential rate of sources with parameters $\bm{\theta}_d$ per unit detector-frame time $t_d$ given the population parameters $\bm{\Lambda}$, and $p_{\rm det}(\bm{\theta}_d)$ is the probability of detecting a source with parameters $\bm{\theta}_d$. 

In this work, we focus on a subset of parameters describing individual events, namely the detector-frame primary masses $m_{1, d}$, the mass ratio $q = m_2/m_1 <1$ and the luminosity distance $d_L$, so that $\bm{\theta}_d = (m_{1, d}, q, d_L)$. 
We assume the distribution of the other extrinsinc parameters to be uniform and neglect all information from spins, as the latter are found to have minor effects on cosmological measurements at current sensitivities~\cite{Tong:2025xvd}. 
We parametrise the rate as a function of the source-frame parameters $\bm{\theta}_s = (m_{1, s}, q, z)$, where $z$ is redshift of the source, and the detector and source frame masses are related by $m_{1, d} = (1+z) m_{1, s}$. The luminosity $d_L(z)$ also depends on $z$ (and other cosmological parameters). 
Thus
\begin{equation}
\label{eq:rate}
    \frac{\dd \mathcal{N}}{\dd \bm{\theta}_d \dd t_d} (m_{1, d}, q, d_L | \bm{\Lambda}) \propto \psi(z | \bm{\Lambda}) ~p_{\rm pop} (m_{\rm 1, s}, q | \bm{\Lambda}) ~\frac{\dd V_c}{\dd z} \frac{1}{1+z} \left| \frac{\partial \bm{\theta}_d}{\partial \bm{\theta}_s} \right|^{-1},
\end{equation}
where $\psi(z | \bm{\Lambda}) \propto \frac{\dd \mathcal{N}}{\dd V_c \dd t_s}$ is the rate of events per comoving volume\footnote{Assuming that the rate is independent of the source sky localisation, which we integrate over, the change of variables from $d_L$ to $V_c$ leads to a trivial overall factor of $4\pi$.} in the source frame, describing the overall evolution of the merger rate with redshift (with $\psi( 0| \bm{\Lambda}) = 1$), and $\left| \frac{\partial \bm{\theta}_d}{\partial \bm{\theta}_s} \right|$ the Jacobian from detector- to source-frame. 
Finally,
\begin{equation}
\label{eq:p_pop}
    p_{\rm pop} (m_{\rm 1, s}, q | \bm{\Lambda}) = p(m_{\rm 1, s} | \bm{\Lambda}) ~p(q | \bm{\Lambda})
\end{equation}
is the probability density function modeling the astrophysical distribution of GW sources, with $p(m_{\rm 1, s} | \bm{\Lambda})$ and $p(q | \bm{\Lambda})$ being respectively the distributions of primary mass and mass ratio. 
As opposed to \cite{LIGOScientific:2025jau}, we mainly work with the mass ratio distribution instead of the secondary mass distribution, which is taken independent of the primary mass. 
See Sec.~\ref{sec:res:gwtc4:m2param} for further discussions.
The following sections give the parametric forms assumed for $\psi(z | \bm{\Lambda})$, $p(m_{1, s} | \bm{\Lambda})$ and $p(q | \bm{\Lambda})$. 

We evaluate the likelihood~\eqref{eq:hierarchical_likelihood} by \ac{MC} integration using posterior samples from the individual events and so-called \textit{injections} to estimate the detection probability $p_{\rm det}(\bm{\theta})$, as described in~\cite{Mastrogiovanni:2023zbw}. 
For real (simulated) data analyses, we use $3 \times 10^3$ ($10^4$) posterior samples from individual events to compute the integral in the numerator in Eq.~\eqref{eq:hierarchical_likelihood} and $10^6$ ($5 \times 10^7$) injections to compute the selection effects integral in the denominator.
The data used is discussed in Sec.~\ref{sec:methods:data} and numerical stability of \ac{MC} integrals in App.~\ref{app:numerical_stability}.
We use the \icarogw~\cite{Mastrogiovanni:2023zbw} \textsc{Python} package's numerical implementation of the hierarchical likelihood, together with the \nessai~\cite{nessai} sampler --- part of the \bilby~\cite{Ashton:2018jfp, Romero-Shaw:2020owr, Smith:2019ucc} Bayesian inference package --- which implements a machine learning-augmented version of the nested sampling algorithm~\cite{Skilling2006NestedSF} making use of normalising flows. 
All analyses are carried out with 2000 live points and a stopping criterion \texttt{dlogZ < 0.1}. Our analyses are produced with the \texttt{icaroverse}~\cite{icaroverse} package, which provides an end-to-end workflow for hierarchical inference with \icarogw as well as mock data simulation.

\subsection{Population models}
\label{sec:methods:pop}

\begin{figure}[tb!]
    \centering
    \includegraphics[width=\linewidth]{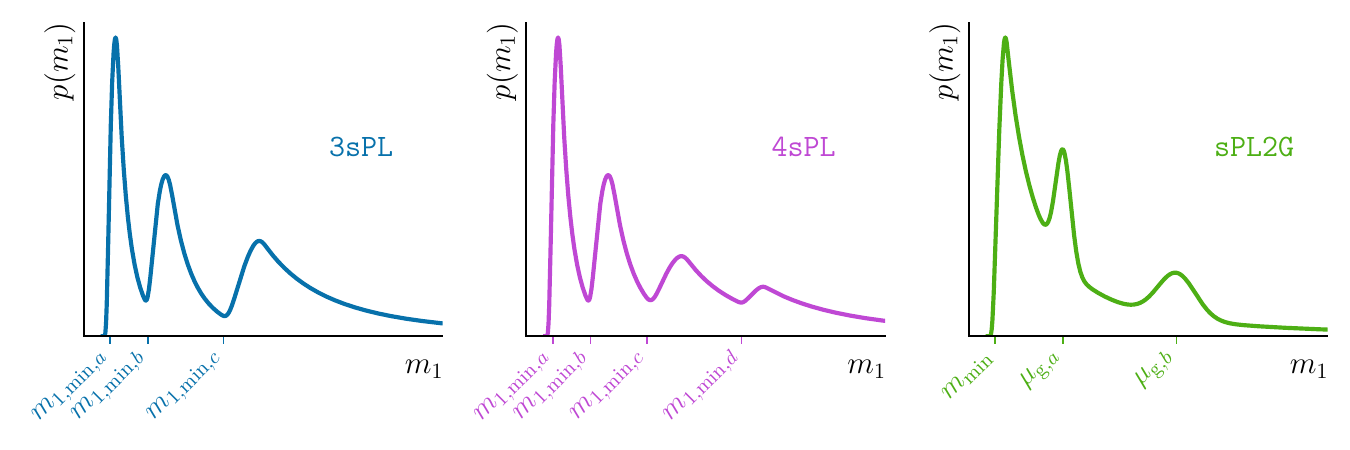}
    \caption{Sketches of the approximate shape of the three different source-frame primary mass distribution parametrisations considered in this work. In blue (left) and purple (center) the new models --- respectively \camelthrees and \camelfours  --- we introduce. In green (right), the fiducial model \sPLGG for comparison. The main mass scales of each model are indicated on the horizontal axes.}
    \label{fig:mass_models_sketches}
\end{figure}

The modelling of the primary mass distribution is crucial for cosmological inference with spectral sirens.
Building on previous studies~\cite{Gennari:2025nho, Bertheas:2025mzd}, we propose two new mass models able to simultaneously capture both low- and high-mass structure in the astrophysical distribution. We then compare these to a fiducial model (see Fig.~\ref{fig:mass_models_sketches} for corresponding sketches). In detail:
\begin{itemize}
    \item The two new models, \texttt{Three} and \texttt{Four smooth Power-Laws} (respectively \camelthrees and \camelfours in the following) are linear combinations of respectively 3 and 4 decreasing truncated power-laws tapered at their lower ends. These models (in particular \camelfours) achieve a good balance between flexibility and simplicity, offering sufficient complexity while remaining fully parametric. We allow the slopes of the power-law components in \camelthrees and \camelfours to have large values (typically up to 200) in order to fit potential sharp features in the mass distribution. This enables us to refine cosmological inference in a way that is not possible with other current parametric models.
    \item For comparison, the fiducial \texttt{Smooth Power-Law + 2 Gaussians} (\sPLGG in the following) models the primary mass distribution as a linear combination of a decreasing truncated power-law and two Gaussian overdensities. The distribution is regularised at its lower end by a smooth tapering function. Compared to power-law components in \camelthrees and \camelfours, the power-law component in \sPLGG is restricted to moderate slope values (typically only up to 12). \sPLGG has the same primary mass parametrisation as the \textsc{MultiPeak} model in~\cite{LIGOScientific:2025jau}, although we have renamed it here for consistency of notation. Notice that in this work it is paired with a parametrisation of the mass ratio's distribution instead of the secondary mass's as done in~\cite{LIGOScientific:2025jau} (see further discussions in Sec.~\ref{sec:res:gwtc4:mass_distrib}).
\end{itemize}
Finally, 
$p(q | \bm{\Lambda})$ is parametrised by a truncated Gaussian in the $[0, 1]$ range (see Eq.~\eqref{eq:model_q}), and $\psi(z | \bm{\Lambda})$ by a Madau-Dickinson function inspired by the star formation rate introduced in~\cite{Madau:2014bja} (see Eq.~\eqref{eq:madau_diskinson_SFR}).  The distributions \camelthrees, \camelfours and \sPLGG  depend respectively on 14, 19 and 10 parameters, in addition to 2 parameters for $p(q)$, 3 parameters for $\psi(z)$ as well as a set of cosmological parameters depending on the cosmological model (see Sec.~\ref{sec:methods:cosmo}). A detailed description of the functional form of our population models, their parameters and corresponding priors is given in App.~\ref{app:models_priors} and tables therein.

\subsection{Cosmological models}
\label{sec:methods:cosmo}

This section highlights different cosmological models considered. Assuming \ac{GR}, with dominant cold dark matter (CDM) and \ac{DE} (relevant for late-universe), the luminosity distance and redshift $z$ of a source are related by
\begin{equation}
    d_L(z) = \frac{c(1+z)}{H_0} \int_0^z \frac{\dd z'}{\left[ \Omega_m (1+z')^3 + (1 - \Omega_m) f(z') \right]^{1/2}}
    \label{eq:dL}
\end{equation}
with
\begin{equation}
    f(z) = \exp \left( 3 \int_0^z \frac{1 + w(z')}{1 + z'} \dd z' \right)
\end{equation}
where $H_0$ is the Hubble constant, $\Omega_m$ the CDM density fraction at present time, and $w(z)$ is the \ac{DE} equation of state parameter.
The first model we consider is the standard flat \LCDM model \cite{LIGOScientific:2025jau} (\texttt{FlatLCDM} in the following), for which $w(z) = - 1$ describing a cosmological constant.
The second model still assumes GR but aims at describing a potential evolution of the \ac{DE} component. We adopt the widely used \ac{CPL} parametrization for the \ac{DE} equation of state~\cite{Chevallier:2000qy, Linder:2002et}:
\begin{equation}
    w(z) = w_0 + w_a \frac{z}{1+z}
    \label{eq:wofz}
\end{equation}
which we refer to as the \texttt{w0waCDM} model. Notice that the subcase $w_a = 0$ (\texttt{wCDM} in the following) corresponds to a constant \ac{DE} equation of state which does not necessarily coincide with a cosmological constant. 

In beyond-\ac{GR} theories, particularly those which aim to solve the cosmological constant problem by modifying gravity on large scales, the \ac{GW} luminosity distance can differ from the \ac{EM} one given in Eq.~\eqref{eq:dL} (and denoted by $d_L^{\rm EM}$ in the following). Motivated by many previous analyses \cite{LIGOScientific:2025jau, Leyde:2022orh, LISACosmologyWorkingGroup:2019mwx, Mancarella:2021ecn, Chen:2023wpj, Liu:2023onj, Mangiagli:2023ize, Tagliazucchi:2025ofb}, the third model we consider, denoted \texttt{Xi0}, introduces two positive parameters $\Xi_0$ and $n$, with
\begin{equation}
\label{eq:Xi0}
    \frac{d_L^{\rm GW}}{d_L^{\rm EM}} = \Xi_0 + \frac{1 - \Xi_0}{(1+z)^n}.
\end{equation}
This parametrisation can describe a wide class scalar-tensor beyond-\ac{GR} theories of gravity (including those with $c_{\rm GW}=1$ implicitly assumed here)~\cite{Horndeski:1974wa, Gleyzes:2014rba, LISACosmologyWorkingGroup:2019mwx, Belgacem:2018lbp}. It recovers GR for $\Xi_0 = 1$, in which case $n$ is a redundant parameter, or $n = 0$ (however, note that our priors will always impose $n>0$).
The fourth and last model considered (which we refer to as \texttt{cM}) assumes that the additional beyond-\ac{GR} friction term in the \ac{GW} propagation is proportional to the \ac{DE} energy fraction via a single parameter $c_M$~\cite{Bellini:2015xja, Alonso:2016suf, Mastrogiovanni:2020gua}. Under the additional assumption that the evolution of the \ac{EM} sector follows flat \LCDM dynamics, one can obtain the luminosity distance ratio~\cite{Lagos:2019kds}:
\begin{equation}
\label{eq:cM}
    \frac{d_L^{\rm GW}}{d_L^{\rm EM}} = \exp \left( \frac{c_M}{2 (1 - \Omega_m)} \ln \frac{1+z}{\sqrt[3]{\Omega_m (1 + z)^3 + (1 - \Omega_m)}} \right).
\end{equation}
This model recovers GR for $c_M = 0$. Prior distributions assumed on the corresponding parameters during the analyses are summarised in Tab.~\ref{tab:priors_cosmo} of the App.~\ref{app:models_priors}.

\subsection{Data}
\label{sec:methods:data}

This section describes the datasets used in our analysis both for real observations (\ac{GWTC}) and mock simulated ones (O5).

\subsubsection{GWTC-4.0}
\label{sec:methods:data:real}

We analyse the data from the latest released \ac{GWTC} catalog, namely the \ac{GW} transients detected up to and including the first part of the fourth observing run (O4a) of the LVK network~\cite{LIGOScientific:2025slb}.

Only \ac{BBH} events are considered, and we impose a threshold \ac{FAR} $< 1 \rm yr^{-1}$ for detection (as in \cite{LIGOScientific:2025jau}, no additional \ac{SNR} cut is imposed). 
Following \cite{LIGOScientific:2025jau} the high mass event GW231123~\cite{LIGOScientific:2025rsn} is excluded due to its properties being more dependent on the choice of waveform model than other events. 
Furthermore, building on previous analyses \cite{Gennari:2025nho, Bertheas:2025mzd} we exclude GW190521~\cite{LIGOScientific:2020iuh, LIGOScientific:2020ufj} and GW190412~\cite{LIGOScientific:2020stg}, respectively very high mass and low mass ratio outliers from the distribution of the rest of events. 
In App.~\ref{app:far_comparison} we check, however, that their inclusion has no influence on our results. Our final dataset contains 150 \ac{BBH} events. 
For comparison, Ref.~\cite{LIGOScientific:2025jau} imposed a \ac{FAR} < $0.25 \rm yr^{-1}$ in its \ac{BBH}-only analyses, for a total of 137 \acp{BBH} events, 13 less than this work.
Notice that~\cite{LIGOScientific:2025jau} also performs analyses including all \acp{CBC} events (\acp{BBH} as well as NS-containing binaries), which we report in some of the following sections (in particular Sec.~\ref{sec:res:gwtc4:MG}) for reference.

To estimate selection effects, we use the injection sets publicly released by the \ac{LVK} collaboration \cite{Essick:2025zed}, which contain \textit{real noise} injections at O3 and O4a sensitivity, combined with semi-analytic injections for O1 and O2. These are the same as the injections used to produce \ac{LVK} population and cosmology analyses on \ac{GWTC} data \cite{LIGOScientific:2025pvj, LIGOScientific:2025jau}.

\subsubsection{Mock O5 catalog}
\label{sec:methods:data:mock}

To provide prospects for the upcoming \ac{LVK} observing run O5, we simulate a mock dataset at O5 sensitivity, considering a universe with \LCDM Planck 2015 best fit cosmology~\cite{Planck:2015fie}. Our mock \ac{BBH} events are drawn from astrophysical rates following a \camelthrees model -- slightly preferred by Bayesian evidence over \sPLGG on current data, as explained in Sec.~\ref{sec:res:gwtc4} -- with parameters fixed at their median values from an analysis with fixed \LCDM Planck 2015 cosmology (see tables of App.~\ref{app:models_priors}). This choice complements other forecasts in the literature \cite{Mancarella:2021ecn, Leyde:2022orh, Borghi:2023opd, Bertheas:2025mzd} assuming simpler models more similar to the \sPLGG.

A number of similar analyses in the literature use Gaussian approximations for the \ac{SNR} and posterior distribution of individual events' parameters \cite{Mastrogiovanni:2022hil, Leyde:2022orh}. While these solutions allow for fast \ac{SNR} computation and generation of posterior samples, they are usually considered valid only in the high-\ac{SNR} limit~\cite{Iacovelli:2022bbs}. 
Here, we adopt a more realistic framework for both detection and parameter estimation, similar to~\cite{Agarwal:2024hld}. Compared to their work, we extend the catalog size by one order of magnitude by considering O5 data.

We consider a four detectors network consisting of the two LIGO~\cite{LIGOScientific:2014pky}, Virgo~\cite{VIRGO:2014yos} and KAGRA~\cite{KAGRA:2020tym}, all operating at 70\% duty cycle with design A+ sensitivity for O5 run~\cite{KAGRA:2013rdx}. This corresponds to a \ac{BNS} range of $330 ~\rm Mpc$, $150 ~\rm Mpc$ and $80 ~\rm Mpc$, respectively. We evaluate the detectability of each \ac{GW} event based on its network matched-filter \ac{SNR} (computed using \texttt{bilby}~\cite{Ashton:2018jfp, Romero-Shaw:2020owr, Smith:2019ucc}), setting the detection threshold at $\rm SNR = 12$. For each event, we simulate Gaussian noise in all detectors from the \acp{PSD} at O5 design sensitivities~\cite{KAGRA:2013rdx}, and model the \ac{GW} signal using the \texttt{IMRPhenomXHM} waveform~\cite{Garcia-Quiros:2020qpx}. The three-dimensional spin components are drawn uniformly in magnitude and isotropic in orientation, namely we are agnostic about the spin distribution, which is not included in our population models. For each event, the remaining parameters (inclination angle, polarisation angle, etc.) are similarly drawn from the same distributions as the ones used as priors for single event \ac{PE}, as discussed below and in App.~\ref{app:mock_data_production:pe}.
Our mock catalog thus contains 1500 detected events spanning a redshift range up to $z \sim 3$, corresponding to $\sim 3$ yrs of observation at O5 sensitivity. 

To make our analysis as realistic as possible, we perform full Bayesian \ac{PE} on the detected events using \texttt{bilby}, to obtain the single event posterior distributions required for hierarchical inference. The priors on the single event parameters used for \ac{PE} are further discussed in App.~\ref{app:mock_data_production:pe}. 
%
We account for selection effects in the analysis by generating \textit{injections} which cover a sufficiently large portion of the parameter space, and assess their detectability with the same procedure used for the mock dataset. Details about the injections can be found in App.~\ref{app:mock_data_production:inj}.

\section{Results}
\label{sec:res}

This section presents our results, focusing on constraints on cosmological parameters from \ac{GWTC} in Sec.~\ref{sec:res:gwtc4}, and forecasts with O5 simulated data in Sec.~\ref{sec:res:O5}. Unless stated otherwise, measured parameters are always quoted as their median value accompanied with their symmetric 68\% \acp{CI}.

\subsection{GWTC-4.0 constraints on cosmology}
\label{sec:res:gwtc4}
We first discuss the differences between the models in terms of Bayesian evidence.
\acp{BF} compared to \camelthrees with a \texttt{FlatLCDM} cosmology are summarised in Tabs. \ref{tab:gwtc4_constraints:GR} and \ref{tab:gwtc4_constraints:MG} for \ac{GR} (\texttt{FlatLCDM} and \texttt{w0waCDM}) and modified \ac{GW} propagation (\texttt{Xi0} and \texttt{cM}) models, respectively.
For all cosmological models considered, the \camelthrees provides a slightly better fit to \ac{GWTC} data than both the \camelfours and the fiducial \sPLGG models, with \acp{BF} ranging between $\mathcal{B} \approx 5 \text{ to } 20$, although these values shouldn't be regarded as strongly statistically decisive. 
We note that \camelfours provides comparable evidence to the \sPLGG model, despite the larger number of parameters. 
This can be explained by the fact that the less flexible \sPLGG is unable to resolve all the features in the mass spectrum (see Fig.~\ref{fig:PPD_m1_gwtc4_LCDM} and discussion in Sec.~\ref{sec:res:gwtc4:mass_distrib} below). 
This trend was already observed on GWTC-3 data for $\Lambda$CDM~\cite{Gennari:2025nho, Bertheas:2025mzd}, even though the fiducial model considered there had only one Gaussian component and no smoothing was applied to the power-law components in \camelthrees. 
We emphasize here that \acp{BF} --- although widely used in the literature as a simple model comparison metric --- may not represent an exhaustive figure of merit for model selection when there are significant regions of the parameter space excluded by numerical stability cuts (see discussion in App.~\ref{app:numerical_stability}).

\subsubsection{BBHs mass distribution}
\label{sec:res:gwtc4:mass_distrib}

Although our primary focus here is the \acp{BBH} population reconstruction for cosmological inference, this measurement also contains crucial information about the astrophysical processes underlying our observations. We briefly discuss below possible physical interpretations of the distributions inferred with the \camelthrees and \camelfours models.

\begin{figure}[!t]
    \centering
    \includegraphics[width=0.8\linewidth]{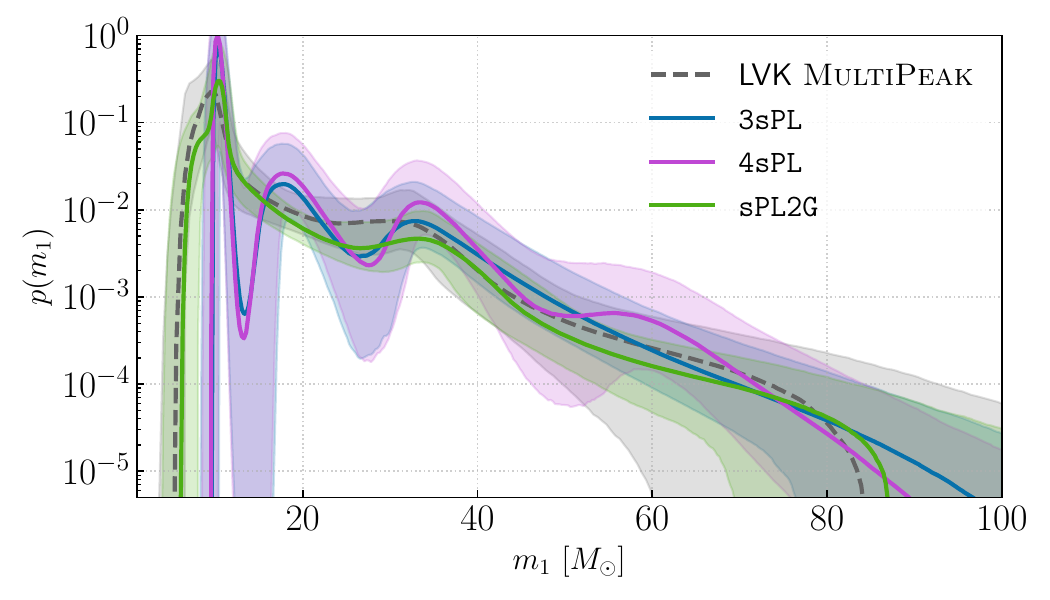}
    \caption[GWTC-4 - PPDs]{Marginal PPD for the \sPLGG (green), \camelthrees (blue) and \camelfours (purple) population models, from analyses inferring $H_0$ and $\Omega_m$ parameters of a \texttt{FlatLCDM} cosmological model. The thick lines denote the median value, while the shaded areas cover the 90\% \ac{CI}. For comparison, the gray curve shows the \ac{LVK} analysis using the \textsc{MultiPeak} model~\cite{LIGOScientific:2025jau}.}
    \label{fig:PPD_m1_gwtc4_LCDM}
\end{figure}

Fig.~\ref{fig:PPD_m1_gwtc4_LCDM} shows the primary mass \acp{PPD}, highlighting the additional structure in the mass spectrum captured by \camelthrees and \camelfours models (blue and purple curves respectively) compared to \sPLGG (green curve), all being nonetheless statistically compatible at 95\% \ac{CI}. 
While the presence of a global $10 \,M_\odot$ maximum has been robustly established since the first catalogs of GW transient events~\cite{LIGOScientific:2020kqk}, the \camelthrees and \camelfours models yield a sharper reconstruction of this feature, in agreement with previous studies~\cite{Toubiana:2023egi, Gennari:2025nho, Bertheas:2025mzd} and recent \ac{GWTC} analyses~\cite{LIGOScientific:2025pvj, LIGOScientific:2025jau, Legred:2026oiz}, including underdensities around $\sim 15 \,M_\odot$ and $\sim 30 \,M_\odot$ and an additional overdensity at $\sim 20 \,M_\odot$.
Several astrophysical formation channels have been proposed to explain the accumulation and dearth of \ac{BBH} mergers around these mass scales. For example, isolated binary-star evolution may account for the observed low-mass features~\cite{Belczynski:2017gds, Giacobbo:2018etu, Neijssel:2019irh}, with mass transfer potentially producing a bimodal structure~\cite{Schneider:2023mxe, Galaudage:2024meo, Legred:2026oiz}.
The \ac{PISN} process is a widely discussed explanation for the $\sim 35 \,M_\odot$ overdensity~\cite{Stevenson:2015bqa, Marchant:2018kun, Woosley:2021xba}, as highlighted in multiple observational studies~\cite{Talbot:2018cva, Karathanasis:2022rtr, Tong:2025wpz, Mould:2026sww}. This interpretation is consistent with our results with the \camelthrees and \camelfours models which recover a steeper fall-off of this component relative to the \sPLGG, with power-law index of $\sim 6$ and $\sim 13$, compared to $\sim 3$.
The \camelfours model also recovers a fourth feature at $\sim 60 \,M_\odot$, similar to~\cite{Pierra:2026ffj, MaganaHernandez:2024qkz}, possibly indicating repeated mergers populating the \ac{PISN} mass gap~\cite{Ginat:2026awh}. More generally, hierarchical mergers could explain the successive overdensities observed in the mass spectrum~\cite{Fragione:2023kqv, Tiwari:2025oah}, although such structure may also arise from a mixture of multiple components~\cite{Colloms:2025hib}. Finally, we emphasize that \ac{BBH} formation channels remain an open problem, given the large uncertainties in both observations and population synthesis models.

The dashed line in Fig.~\ref{fig:PPD_m1_gwtc4_LCDM} shows the LVK \textsc{MultiPeak} results~\cite{LIGOScientific:2025jau}, which can be compared with the \sPLGG curve (modulo the setup differences mentioned in Sec.~\ref{sec:methods:pop} above). While both show similar features,
the small variations --- in particular a marginal shift of the position of features --- may be attributed to our parametrisation in terms of the mass ratio, rather than the secondary mass as in~\cite{LIGOScientific:2025jau}, as discussed in Sec.~\ref{sec:res:gwtc4:m2param}.
Finally, we observed that the \acp{PPD} are essentially the same for the different cosmological models, therefore we only report those for \texttt{FlatLCDM} analyses.

\subsubsection{\texorpdfstring{\LCDM}{LCDM}}
\label{sec:res:gwtc4:LCDM}

\begin{figure}[!b]
    \centering
    \includegraphics[width=0.75\linewidth]{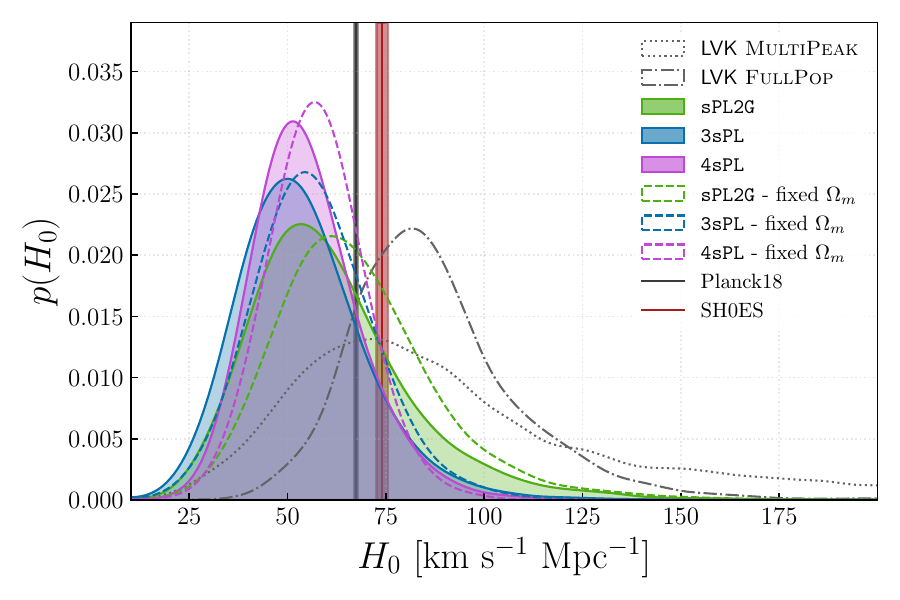}
    \caption[GWTC-4 - \LCDM - $H_0$ posterior]{\ac{GWTC} constraints on the Hubble constant $H_0$ with the three population models \sPLGG (green), \camelthrees (blue), \camelfours (purple) considered here. Solid (resp.~dashed) curves denote marginalised posterior distribution on $H_0$ for analyses where $\Omega_m$ is inferred (resp.~fixed). For comparison, the grey dashed (resp.~dash-dotted) line represents the \ac{LVK} result \cite{LIGOScientific:2025jau} with the \textsc{MultiPeak} (resp.~\textsc{FullPop} and all CBC events) model and fixed $\Omega_m$. 
    The black and red shaded areas identify the 68\% \ac{CI}~constraints on $H_0$ inferred from the CMB anisotropies by Planck~\cite{Planck:2018vyg} and in the local Universe by SH0ES~\cite{Riess:2021jrx}.}
    \label{fig:gwtc4_LCDM_H0_posterior}
\end{figure}

In the context of the standard $\Lambda$CDM cosmology, the best marginalised constraints on the Hubble constant are obtained with the \camelfours model and are $H_0 = 53.3^{+14.0}_{-10.8} ~\rm km \,s^{-1} \,Mpc^{-1}$, while the \camelthrees model yields $H_0 = 52.1^{+16.1}_{-13.3} ~\rm km \,s^{-1} \,Mpc^{-1}$ and \sPLGG yields $H_0 = 60.8^{+24.4}_{-16.7} ~\rm km \,s^{-1} \,Mpc^{-1}$ (see Tab.~\ref{tab:gwtc4_constraints:GR}). 
As can be seen in Fig.~\ref{fig:gwtc4_LCDM_H0_posterior} and Tab.~\ref{tab:gwtc4_constraints:GR}, results only mildly improve when fixing $\Omega_m$ during the inference.
The posterior distributions are shown in Fig.~\ref{fig:gwtc4_LCDM_H0_posterior}, corresponding to a precision\footnote{The \emph{precision} is defined here as half of the width of the $68\%$ confidence interval divided by the median value.} on the $H_0$ measurement of 23\%, 28\% and 34\% respectively.

\begin{table}
    \centering
    \begin{adjustbox}{center}
    \begin{tabular}{@{\extracolsep\fill}llccccc}
        \toprule%
        \multicolumn{2}{@{}r@{}}{\textbf{Parameters $\rightarrow$}} & \multirow{3}{*}{\makecell[c]{$H_0$\\$\rm [km \,s^{-1} \,Mpc^{-1}]$}} & \multirow{3}{*}{$\Omega_m$} & \multirow{3}{*}{$w_0$} & \multirow{3}{*}{$w_a$} & \multirow{3}{*}{$\log_{10} \mathcal{B}$} \\ 
        \textbf{$\downarrow$ Analyses} & & & & & & \\ \cmidrule{1-2} 
        Cosmology & Population & & & & & \\
        \midrule
        \multirow{3}{*}{\texttt{FlatLCDM}} & \camelthrees & $52.1^{+16.1(30.2)}_{-13.3(20.9)}$ & $0.46^{+0.34(0.48)}_{-0.31(0.41)}$ & - & - & $0$ \\
        & \camelfours & $53.3^{+14.0(25.7)}_{-10.8(17.3)}$ & $0.51^{+0.31(0.43)}_{-0.30(0.43)}$ & - & - & $-0.6$ \\
        & \sPLGG & $57.9^{+21.3(42.6)}_{-15.4(23.9)}$ & $0.49^{+0.33(0.45)}_{-0.31(0.42)}$ & - & - & $-0.8$ \\
        \midrule
        \multirow{3}{*}{\makecell[l]{\texttt{FlatLCDM} \\(fix $\Omega_m$)}} & \camelthrees & $55.8^{+15.3(26.9)}_{-13.4(21.7)}$ & - & - & - & $+0.2$ \\
        & \camelfours & $57.1^{+11.5(19.9)}_{-11.0(18.3)}$ & - & - & - & $-0.6$ \\
        & \sPLGG & $64.9^{+20.7(40.5)}_{-16.4(25.8)}$ & - & - & - & $-0.7$ \\
        \midrule
        \multirow{3}{*}{\texttt{w0waCDM}} & \camelthrees & $54.6^{+25.2(55.2)}_{-14.9(22.5)}$ & $0.43^{+0.36(0.50)}_{-0.29(0.38)}$ & $-1.6^{+1.3(2.0)}_{-1.0(1.3)}$ & $-1.0^{+1.6(2.4)}_{-1.3(1.8)}$ & $-0.1$ \\
        & \camelfours & $51.2^{+14.1(26.5)}_{-10.3(16.2)}$ & $0.59^{+0.27(0.37)}_{-0.29(0.44)}$ & $-1.4^{+1.3(2.0)}_{-1.0(1.4)}$ & $-1.1^{+1.5(2.3)}_{-1.2(1.7)}$ & $-0.6$ \\
        & \sPLGG & $59.6^{+29.2(62.3)}_{-17.4(26.1)}$ & $0.49^{+0.33(0.45)}_{-0.31(0.42)}$ & $-1.5^{+1.4(2.1)}_{-1.0(1.3)}$ & $-1.0^{+1.6(2.4)}_{-1.3(1.8)}$ & $-0.8$ \\
        \bottomrule
    \end{tabular}
    \end{adjustbox}
    \caption[GWTC-4 - Constraints]{Constraints from {GWTC-4.0} on cosmological parameters from all analyses with \ac{GR} models. Quoted as $\rm median \pm 68\% ~(90\%)$ \ac{CI}. The rightmost column shows the \acp{BF} $\mathcal{B}$ comparing analyses with respect to the \camelthrees + \texttt{FlatLCDM} model.}
    \label{tab:gwtc4_constraints:GR}
\end{table}

For comparison, the LVK collaboration reports constraints of $H_0 = 81.1^{+45.2}_{-26.7} ~\rm km \,s^{-1} \,Mpc^{-1}$ and $H_0 = 84.1^{+22.4}_{-16.2} ~\rm km \,s^{-1} \,Mpc^{-1}$, respectively from spectral sirens analysis using the \textsc{MultiPeak} mass model with \ac{BBH} events only, and from dark sirens analysis using the \textsc{FullPop} model, including galaxy catalogs and NSs~\cite{LIGOScientific:2025jau}.  These correspond to a precision of 44\% and 23\% respectively.
It follows that our \camelfours (\camelthrees) population model improves the $H_0$ constraints by $\sim 50\%$ ($\sim 35\%$) compared to \ac{LVK} results with \ac{BBH} events only, and provides comparable precision to that achieved using the full population with galaxy catalogs.
This improvement is expected due to the additional structure resolved by the power-law models, because the spectral sirens method relies on distinguishing features in order to extract redshift information from the population.

Although our results are statistically consistent with both Planck and SH0ES measurements, our $H_0$ posteriors peak at lower values compared to the Hubble-tension region and the \ac{LVK} results.
Even ignoring potential modelling systematics, variance associated to finite catalogs may lead to similar shifts in the $H_0$ posteriors at current sensitivities, as shown in Ref.~\cite{Mould:2026sww} (see Fig.~8). However,
this shift may also suggest the presence of systematic effects that have not been accounted for.
Several of the approximations adopted in our analysis could, in principle, introduce biases -- for example, inaccurate modeling of individual parameters or neglecting population-level correlations between them~\cite{Mukherjee:2021rtw, Pierra:2023deu, Agarwal:2024hld}.
Our Gaussian model for the mass ratio allows for support at very low secondary masses through the low-mass-ratio tail. However, if this mechanism was driving the result, it would tend to favour higher values of $H_0$, which is the opposite of what we observe. This is further supported by additional analyses in which the secondary mass is parameterised directly (see Sec.~\ref{sec:res:gwtc4:m2param}). 
Another potential source of bias is the relatively restrictive modeling of the primary mass around $10\,M_\odot$. Our \camelthrees/\camelfours parameterisations enforce both a sharp low-mass cutoff and an asymmetric functional form, which could limit the flexibility of the inferred distribution.
However, we see from Fig.~\ref{fig:PPD_m1_gwtc4_LCDM} that the \sPLGG \ac{PPD} shows support below $10\,M_\odot$ although the corresponding $H_0$ posterior peaks at the same place (Fig.~\ref{fig:gwtc4_LCDM_H0_posterior}). Sec.~\ref{sec:res:gwtc4:m2param} presents some additional \camelthrees analyses with variations for the secondary mass / mass ratio parametrisations and for the low mass end of the spectrum. A thorough comparison exploring additional parametric modelling variations as well as more agnostic modelling approaches, designed to test the robustness of these assumptions and assess the potential systematics, will be presented in an upcoming study. 
However, we note that a similar shift in the $H_0$ posterior has been observed using different or more flexible parametrisations~\cite{Mould:2026sww, Tagliazucchi:2026gxn}.

Finally, we have checked that using a different criterion for event selection (i.e. a \ac{FAR} $<0.25 \rm yr^{-1}$ as in~\cite{LIGOScientific:2025jau}) does not affect the results (see App.~\ref{app:far_comparison}). 

\subsubsection{Secondary mass parametrisation}
\label{sec:res:gwtc4:m2param}

\begin{figure}[h!]
    \centering
    \begin{subfigure}[c]{0.7\textwidth}
        \centering
        \includegraphics[width=\linewidth]{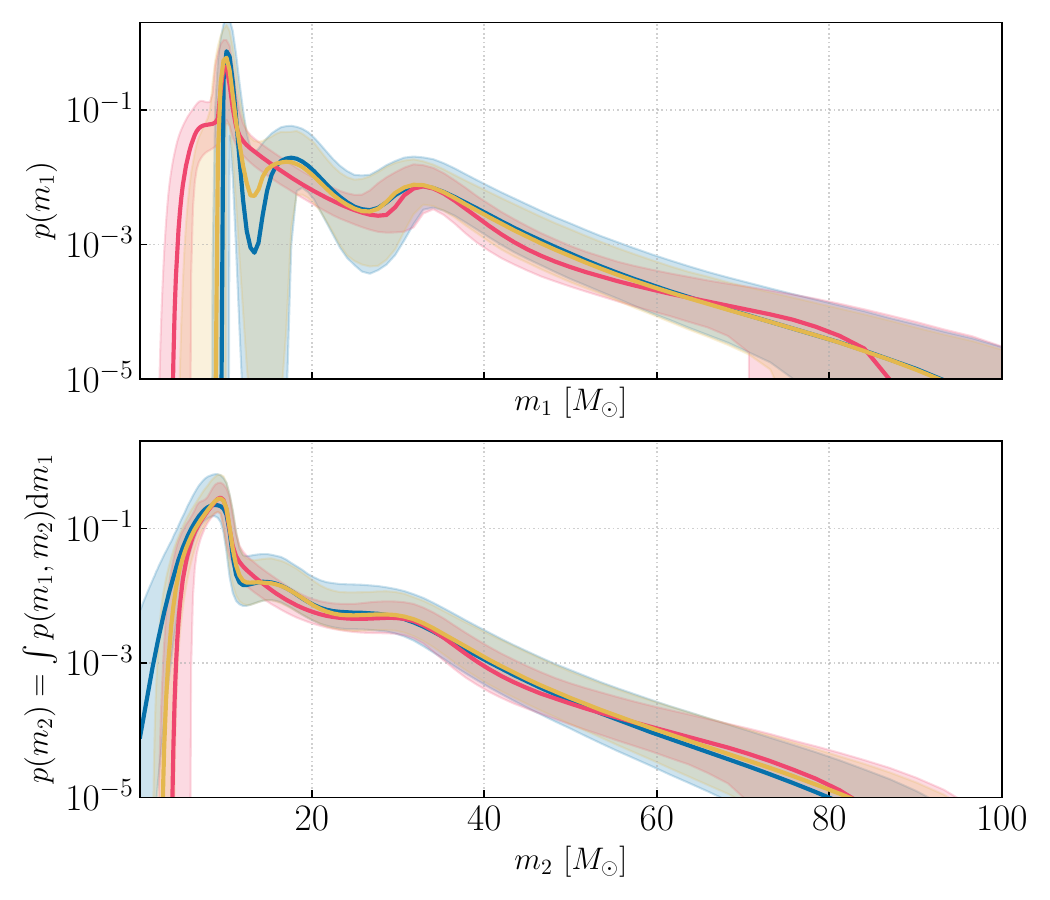}
        \caption{}
        \label{fig:m1m2_param_PPD}
    \end{subfigure}
    \begin{subfigure}[c]{0.6\textwidth}
        \centering
        \includegraphics[width=\linewidth]{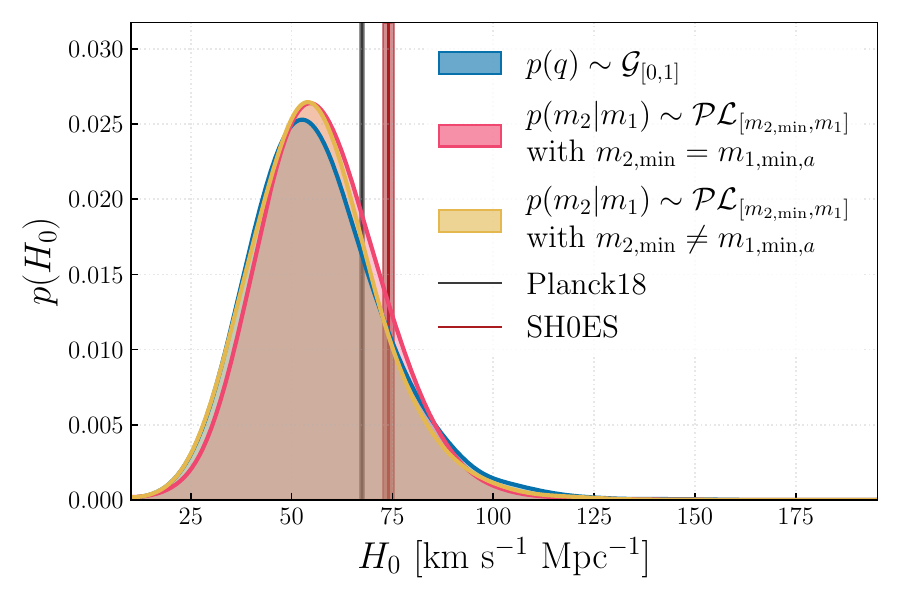}
        \caption{}
        \label{fig:m1m2_param_H0}
    \end{subfigure}
    \caption[GWTC-4 - m2 parametrisations]{
    \textbf{(a)} Primary (top) and secondary (bottom) mass marginal PPD and \textbf{(b)} $H_0$ posterior distribution (from fixed $\Omega_m$ analyses) obtained on GWTC-4.0 data using the \camelthrees primary mass model with three different parametrisations for the secondary mass (see main text for details).
    }
    \label{fig:m1m2_param}
\end{figure}

Motivated by the potential shift observed in the $H_0$ posterior (see Sec.~\ref{sec:res:gwtc4:LCDM}), we investigate the robustness of our modelling choices by testing alternative parametrisations for the secondary mass $m_2$. Instead of the mass ratio $q$, we adopt a power-law distribution for $m_2$ conditioned on $m_1$, i.e. $m_{\rm 2, min} < m_2 < m_1$, with either $m_{\rm 2, min} = m_{{\rm 1, min}, a}$ or $m_{\rm 2, min} < m_{{\rm 1, min}, a}$ (where $m_{{\rm 1, min}, a}$ is the minimum mass truncation of the low-mass power-law component of the \camelthrees model, see App.~\ref{app:models_priors:pop} for further details and Fig.~\ref{fig:mass_models_sketches} for illustration). The former coincides with the secondary-mass parametrisation used in the \ac{LVK} spectral sirens analysis~\cite{LIGOScientific:2025jau}, while the latter represents an intermediate case relative to our $p(q)$ parametrisation. For the primary mass, we adopt a simplified variant of the \camelthrees model in which the three power-law components share a common maximum mass cutoff $m_{\rm 1, max}$. We use slightly broader priors than those reported in App.~\ref{app:models_priors} for $m_{\rm 1, min, a}$, $m_{{\rm 1, min}, b}$ and $m_{{\rm 1, min}, c}$, namely $[1, 15] \,M_\odot$, $[1, 30] \,M_\odot$ and $[1, 50] \,M_\odot$, respectively, with the ordering condition $m_{{\rm 1, min}, a} < m_{{\rm 1, min}, b} < m_{{\rm 1, min}, c}$. In this section, we adopt a \texttt{FlatLCDM} model with fixed $\Omega_m = 0.308$. We have verified that these modifications do not affect the results relative to the modelling assumptions adopted in the rest of the paper.

Fig.~\ref{fig:m1m2_param_PPD} shows the primary (top panel) and secondary (bottom panel) marginal \ac{PPD}s for the cases considered. 
When $m_{\rm 2, min} = m_{{\rm 1, min}, a}$ (pink curve), the \camelthrees model behaves similarly to the \sPLGG, with one power-law component fitting the bulk of the mass spectrum and the other two capturing overdensities at $10\,M_\odot$ and $35\,M_\odot$. 
Compared to the other parametrisations, this scenario recovers support at masses $< 10\,M_\odot$, rather than favouring a sharp truncation of the first feature. 
This occurs because $m_{{\rm 1, min}, a}$ must also accommodate low-$m_2$ events, as $m_1$ and $m_2$ share the same minimum cutoff. 
Physically, this corresponds to modelling a common minimum mass for the full BH distribution, rather than for the primary and secondary components independently, and the resulting low-mass plateau is therefore expected from the ordering $m_2 < m_1$. 
This differs from the other cases, where $m_{{\rm 1, min}, a}$ describes only the minimum primary mass and is thus inferred at higher values than $m_{\rm 2, min}$ when the two components are modelled independently (yellow curve). 
The $q$ parametrisation (blue curve) is similar to the limiting case $m_{\rm 2, min} \to 0$, since $p(q)$ has support down to $q=0$. Despite this approximation, both approaches recover the same primary mass spectrum.

Fig.~\ref{fig:m1m2_param_H0} shows the corresponding $H_0$ posteriors and highlights that the cosmological inference is invariant under different secondary-mass parametrisations. We measure ${H_0 = 55^{+17}_{-14} ~\rm km \,s^{-1} \,Mpc^{-1}}$, ${H_0 = 57^{+16}_{-13} ~\rm km \,s^{-1} \,Mpc^{-1}}$ and ${H_0 = 55^{+17}_{-14} ~\rm km \,s^{-1} \,Mpc^{-1}}$ for the blue, pink and yellow cases, respectively.
This non-trivial result allows us to draw several conclusions. First, a sharp low-mass cutoff of the $10\,M_\odot$ peak does not enhance the cosmological inference. This holds regardless of whether sub-$10\,M_\odot$ support is resolved, consistent with independent findings in Refs.~\cite{Tagliazucchi:2026gxn, Mould:2026sww} using different methods. Second, the resolvability of the $15\,M_\odot$ underdensity and the $20\,M_\odot$ peak carries no additional information on the Hubble constant. Likewise, the $H_0$ measurement is robust to small differences in the $35\,M_\odot$ peak and in the high-mass end of the distribution. Third, low-mass tails in the secondary-mass distribution introduced by modelling the mass ratio with a Gaussian, although visible in the reconstructed $m_2$, do not bias the $H_0$ measurement. A more comprehensive investigation of these findings and their astrophysical implications will be presented in a forthcoming study.

\subsubsection{Dark energy}
\label{sec:res:gwtc4:DE}

\begin{figure}[!b]
    \centering
    \includegraphics[width=0.6\linewidth]{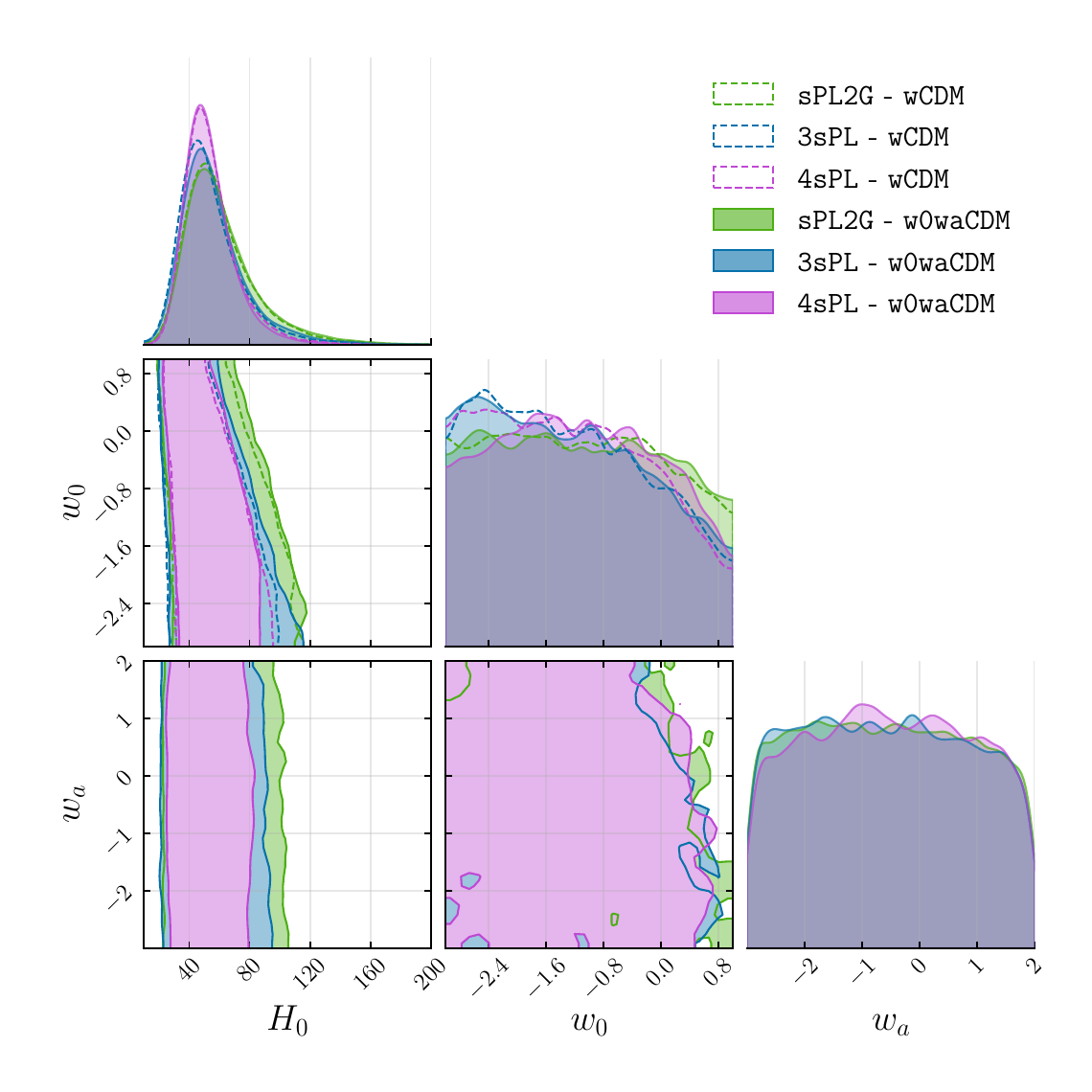}
    \caption[GWTC-4 - \wowaCDM]{\ac{GWTC} constraints on \texttt{wCDM} and \texttt{w0waCDM} cosmological models obtained with the \sPLGG (green), \camelthrees (blue) and \camelfours (purple) population models. Posteriors on $\Omega_m$ are not informative and do not correlate with $w_0$ and $w_a$, hence not displayed. The solid black line denote the \LCDM value of the beyond-\LCDM parameters. Contours indicate 90\% credible area. 
    }
    \label{fig:gwtc4_w0waCDM}
\end{figure}

Assuming the \texttt{w0waCDM} model in Eq.~\eqref{eq:wofz} for dynamical \ac{DE}, our results demonstrate that current \ac{GW} data is unable to constrain $(w_0,w_a)$ together with $H_0$. 
We find that positive values of $w_0$ are unfavoured by data (see the central lower panel of Fig.~\ref{fig:gwtc4_w0waCDM}), although the posterior distribution on $w_0$ and $w_a$ is mostly uninformative, irrespectively of the population model.
The values excluded by the data roughly coincide with unphysical region $w_0 + w_a > 0$, which would prevent the occurrence of an early matter-dominated era, motivating the prior constraint $w_0 + w_a < 0$ adopted in some cosmological analyses~\cite{DESICollaboration2025ddr}. Although we report only results obtained with agnostic priors, we note that imposing such physical constraint does not significantly affect our results, yielding no additional information beyond that imposed by the prior.
Fig.~\ref{fig:gwtc4_w0waCDM} additionally highlights the absence of correlations between DE parameters and $H_0$ at current sensitivities, leading to constraints on the Hubble constant very similar as in the \LCDM case (see Tab.~\ref{tab:gwtc4_constraints:GR}). The same conclusions apply to the case of \texttt{wCDM} (dashed posteriors in Fig.~\ref{fig:gwtc4_w0waCDM}), as assuming a constant \ac{DE} equation of state does not improve the constraints on the $w_0$ parameter. For additional discussion about the limited constraints obtained on \ac{DE} parameters, we refer the reader to Sec.~\ref{sec:res:O5} about O5 forecasts.

\subsubsection{Modified GW propagation}
\label{sec:res:gwtc4:MG}

\begin{table}[b!]
    \centering 
    \begin{adjustbox}{center}
    \begin{tabular}{@{\extracolsep\fill}llccccc}
        \toprule%
        \multicolumn{2}{@{}r@{}}{\textbf{Parameters $\rightarrow$}} & \multirow{3}{*}{\makecell[c]{$H_0$\\$\rm [km \,s^{-1} \,Mpc^{-1}]$}} & \multirow{3}{*}{$\Xi_0$} & \multirow{3}{*}{$c_M$} & \multirow{3}{*}{$\log_{10} \mathcal{B}$} \\ 
        \textbf{$\downarrow$ Analyses} & & & & & \\ \cmidrule{1-2} 
        Cosmology & Population & & & & \\
        \midrule
        \multirow{3}{*}{\makecell[l]{\texttt{Xi0} \\(fix $\Omega_m$)}} & \camelthrees & $71.7^{+27.5(40.6)}_{-23.2(34.9)} $ & $1.8^{+1.7(3.7)}_{-0.7(1.0)}$ & - & $-0.2$ \\
        & \camelfours & $70.0^{+25.8(40.4)}_{-20.5(30.2)}$ & $1.4^{+0.9(2.2)}_{-0.5(0.7)}$ & - & $-1$ \\
        & \sPLGG & $69.4^{+28.9(42.8)}_{-23.3(34.4)}$ & $1.2^{+0.9(1.9)}_{-0.4(0.6)}$ & - & $-1.3$ \\
        \midrule
        \multirow{3}{*}{\makecell[l]{\texttt{Xi0} \\(fix $H_0, \Omega_m$)}} & \camelthrees & - & $1.6^{+1.3(3.1)}_{-0.5(0.7)}$ & - & $0$ \\
        & \camelfours & - & $1.3^{+0.6(1.4)}_{-0.3(0.4)}$ & - & $-1$ \\
        & \sPLGG & - & $1.2^{+0.7(1.7)}_{-0.3(0.5)}$ & - & $-1.1$ \\
        \midrule
        \multirow{3}{*}{\makecell[l]{\texttt{cM} \\(fix $\Omega_m$)}} & \camelthrees & $69.8^{+28.3(42.2)}_{-24.4(36.2)}$ & - & $1.7^{+1.9(2.8)}_{-2.0(3.3)}$ & $-0.2$ \\
        & \camelfours & $64.6^{+27.1(42.8)}_{-21.0(30.8)}$ & - & $0.7^{+1.7(2.8)}_{-1.7(2.7)}$ & $-1.1$ \\
        & \sPLGG & $63.1^{+29.6(45.9)}_{-23.2(33.2)}$ & - & $0.3^{+1.9(3.2)}_{-1.9(3.0)}$ & $-1.3$ \\
        \midrule
        \multirow{3}{*}{\makecell[l]{\texttt{cM} \\(fix $H_0, \Omega_m$)}} & \camelthrees & - & - & $1.4^{+1.5(2.4)}_{-1.2(1.9)}$ & $-0.4$ \\
        & \camelfours & - & - & $0.8^{+1.0(1.9)}_{-0.8(1.2)}$ & $-1$ \\
        & \sPLGG & - & - & $0.4^{+1.5(2.6)}_{-1.2(1.9)}$ & $-1.4$ \\
        \bottomrule
    \end{tabular}
    \end{adjustbox}
    \caption[GWTC-4 - Constraints]{Constraints from {GWTC-4.0} on cosmological parameters from all analyses with modified \ac{GW} propagation models. Quoted as $\rm median \pm 68\% ~(90\%)$ \ac{CI}. The rightmost column show the \acp{BF} $\mathcal{B}$ comparing analyses results on GWTC-4 data. The values are always given as \ac{BF} with respect to the \camelthrees + \texttt{FlatLCDM} analysis (see Tab.~\ref{tab:gwtc4_constraints:GR}).}
    \label{tab:gwtc4_constraints:MG}
\end{table}

\begin{figure}
    \centering
    \begin{subfigure}[c]{0.49\textwidth}
        \centering
        \includegraphics[width=\linewidth]{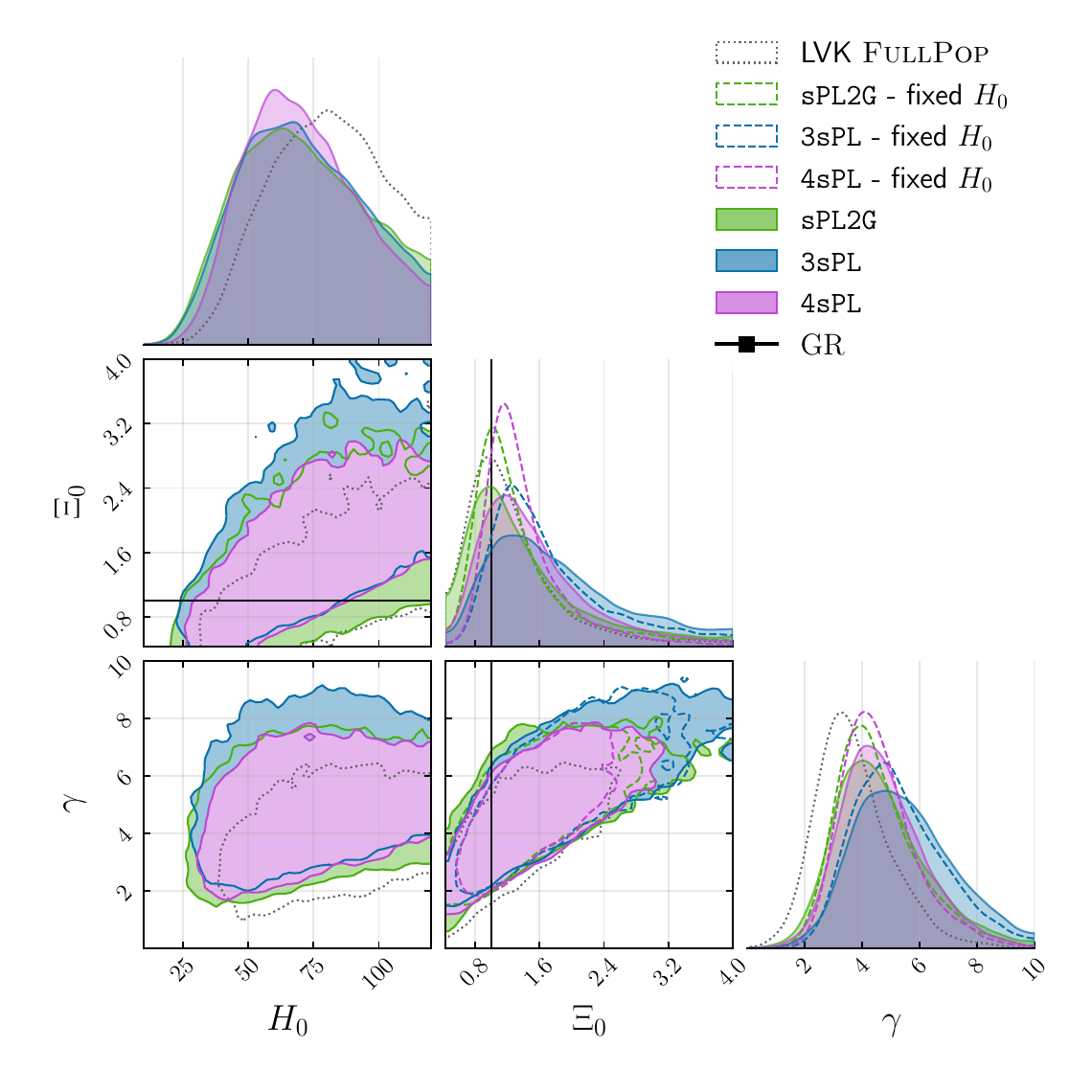}
        \caption{}
        \label{fig:gwtc4_Xi0}
    \end{subfigure}
    \begin{subfigure}[c]{0.49\textwidth}
        \centering
        \includegraphics[width=\linewidth]{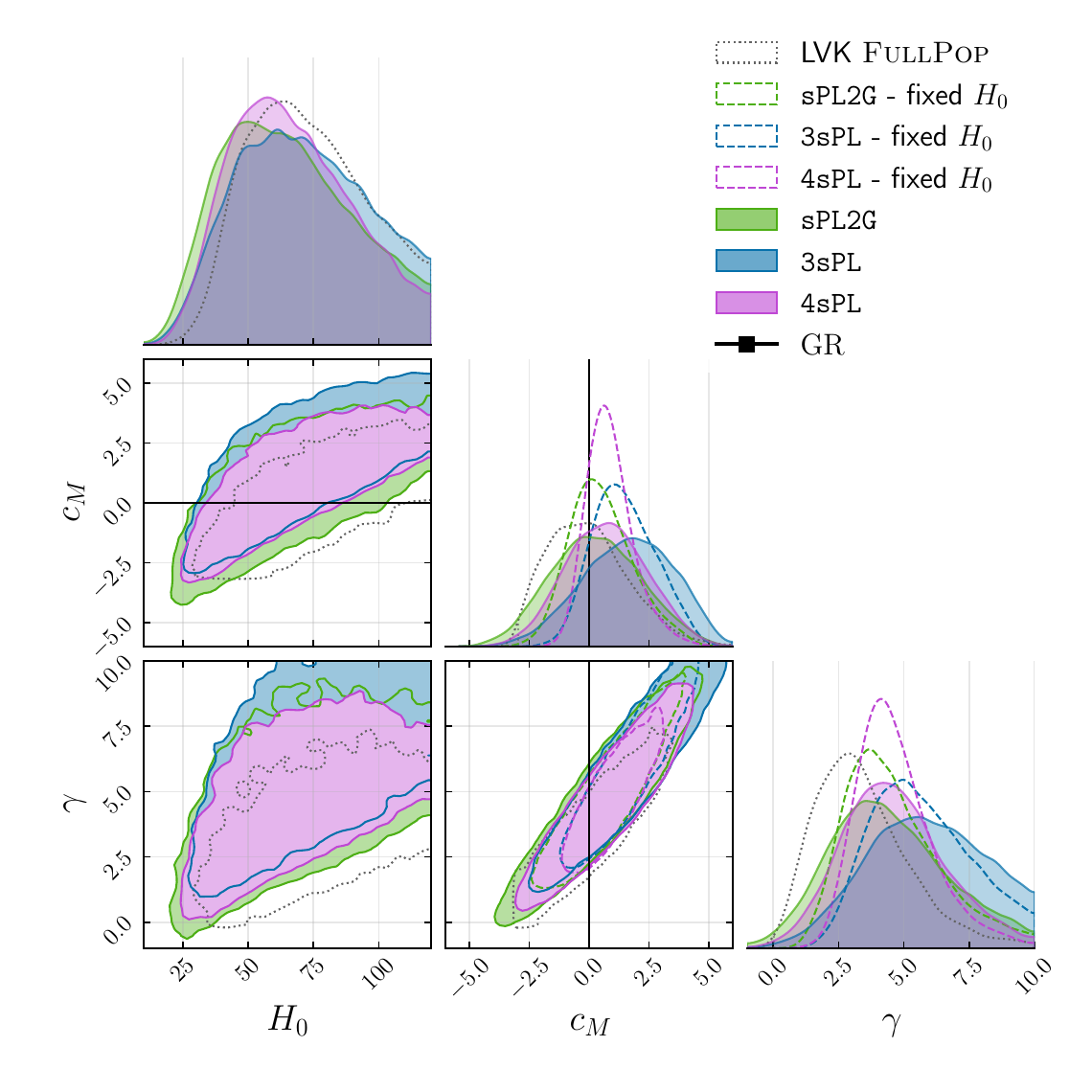}
        \caption{}
        \label{fig:gwtc4_cM}
    \end{subfigure}
    \caption[GWTC-4 - MG]{\ac{GWTC} constraints and correlations on parameters of modified \ac{GW} propagation models \texttt{Xi0} \textbf{(a)} and \texttt{cM} \textbf{(b)}, obtained with the \sPLGG (green), \camelthrees (blue) and \camelfours (purple) population models. Solid (resp.~dashed) lines posteriors denote results of analyses with inferred (resp.~fixed) $H_0$. The solid black line denote the \ac{GR} value of beyond-\ac{GR} parameters. The gray dotted posteriors denote \ac{LVK} results from \cite{LIGOScientific:2025jau} with the \textsc{FullPop} model (see text). Contours indicate 90\% credible area.}
    \label{fig:gwtc4_beyond-GR}
\end{figure}

With the beyond-\ac{GR} models given in Eqs.~\eqref{eq:Xi0} and \eqref{eq:cM}, we obtain marginal constraints of $\Xi_0 = 1.4^{+0.9}_{-0.5}$ and $c_M = 0.7^{+1.7}_{-1.7}$ with the \camelfours model (see Fig.~\ref{fig:gwtc4_beyond-GR} and Tab.~\ref{tab:gwtc4_constraints:MG} for results with other models as well), similar to those reported by \ac{LVK}~\cite{LIGOScientific:2025jau} including NSs and galaxy catalogs, which finds $\Xi_0 = 1.2^{+0.8}_{-0.4}$ and $c_M = 0.3^{+1.6}
_{-1.4}$.
Our constraints are entirely compatible with the GR values of the $\Xi_0$ and $c_M$ parameters (see solid black lines in Fig.~\ref{fig:gwtc4_beyond-GR}).
Note that, in order to limit computational costs, we cap the number of inferred parameters by fixing $\Omega_m = 0.308$, having checked that it does not correlate with the parameters of interest in this section.

The central lower frames of Figs.~\ref{fig:gwtc4_Xi0} and \ref{fig:gwtc4_cM} highlight the expected correlation between modified \ac{GW} propagation parameters and $\gamma$, which describe the overall evolution of the merger rate with redshift. 
These correlations, first discussed in \cite{Leyde:2022orh} and also observed in \cite{LIGOScientific:2025jau}, are due to $\Xi_0$ and $c_M$ affecting the observed space-time volume, and therefore the expected number of events. 
The left central frames also show the strong correlation of $\Xi_0$ and $c_M$ with $H_0$, as expected from degeneracies in the $d_L^{\rm GW}(z)$ parametrisations in the redshift range spanned by the data. 
If we fix $H_0$ to the Planck value,
the constraints on beyond-\ac{GR} parameters become significantly more precise, down to $\Xi_0 = 1.3^{+0.6}_{-0.3}$ and $c_M = 0.8^{+1.0}_{-0.8}$. The results are comparable to the \ac{LVK} analysis using narrow priors on $H_0$, which finds $\Xi_0 = 1.0^{+0.4}_{-0.2}$ and $c_M = 0.3^{+1.4}_{-1.0}$.

While for \ac{GR} models the best constraints on $H_0$ are achieved (in descending order) by the \camelfours, \camelthrees and \sPLGG models, a different ordering is observed for beyond-\ac{GR} models, where now \camelthrees ranks third (see Fig.~\ref{fig:gwtc4_beyond-GR} and Tab.~\ref{tab:gwtc4_constraints:MG}).
This result applies to both $H_0$ and beyond \ac{GR} parameters, due to their strong correlation.
To explain this, first recall that for $z \ll 1$ the $d_L - z$ relation is mainly affected by $H_0$, while for $z \gtrsim 1$ it is affected by both $H_0$, and $\Xi_0$ or $c_M$. 
Now consider the population distribution. 
Close-by detectable GW events can be sourced by the whole spectrum of masses, so if this contains sharp features one expects improved $H_0$ constraints, which is precisely the effect observed in Sec.~\ref{sec:res:gwtc4:LCDM}: models recovering sharper features (\camelthrees and \camelfours) yield better $H_0$ constraints. 
At $z \gtrsim 1$ on the other hand, only the loudest high mass events are detectable. 
As a result, one expects better constraints on $\Xi_0$ or $c_M$ in the presence of features at high mass in the \ac{BBH} distribution, which is the observed effect in Fig.~\ref{fig:gwtc4_beyond-GR}: models capturing high mass features (\camelfours and \sPLGG) provide better constraints than models which do not (\camelthrees), as anticipated in Sec.~\ref{sec:res:gwtc4}.

\subsection{Future O5 prospects}
\label{sec:res:O5}
This section presents O5 forecasts for constraints on cosmological parameters using spectral sirens. The generation and properties of the mock O5 catalogue have been discussed in Sec.~\ref{sec:methods:data:mock}; the dataset contains 1500 detected events drawn from the preferred \camelthrees model assuming a \texttt{FlatLCDM} cosmology with Planck values~\cite{Riess:2021jrx}.
The simulated data is then analysed with the different cosmological models presented in Sec.~\ref{sec:methods:cosmo}, using the \camelthrees model for the primary mass distribution.
In all cases, the posterior distributions on cosmological parameters are  compatible (at the 68\% confidence level) with their injected value.

\begin{figure}[!b]
    \centering
    \begin{subfigure}[c]{0.45\textwidth}
        \centering
        \includegraphics[width=\linewidth]{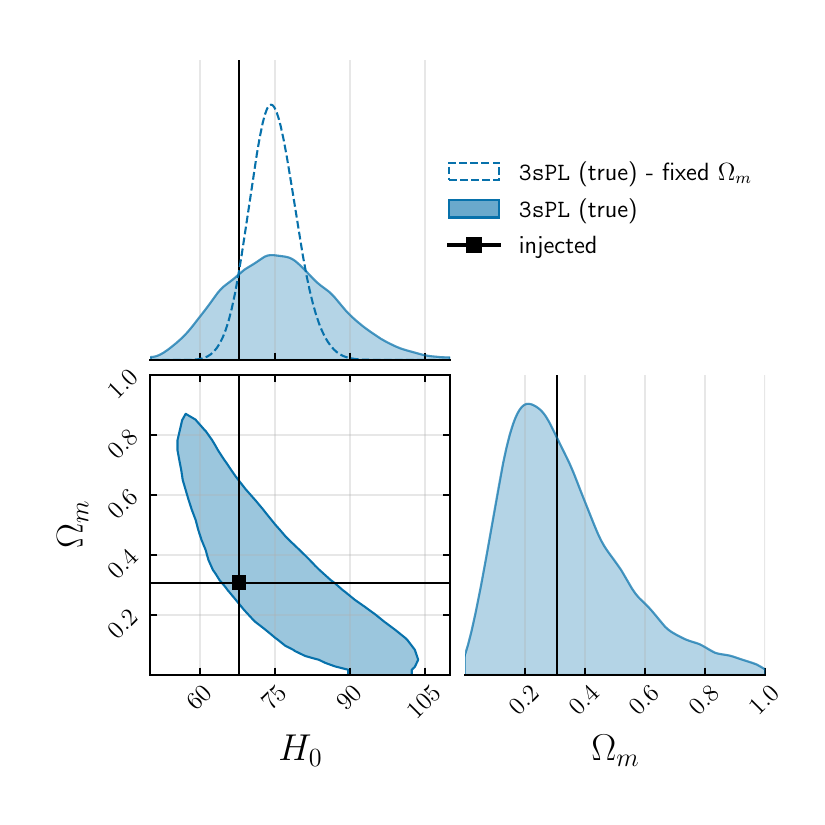}
        \caption{}
        \label{fig:O5_LCDM_corner}
    \end{subfigure}
    \begin{subfigure}[c]{0.54\textwidth}
        \centering
        \includegraphics[width=\linewidth]{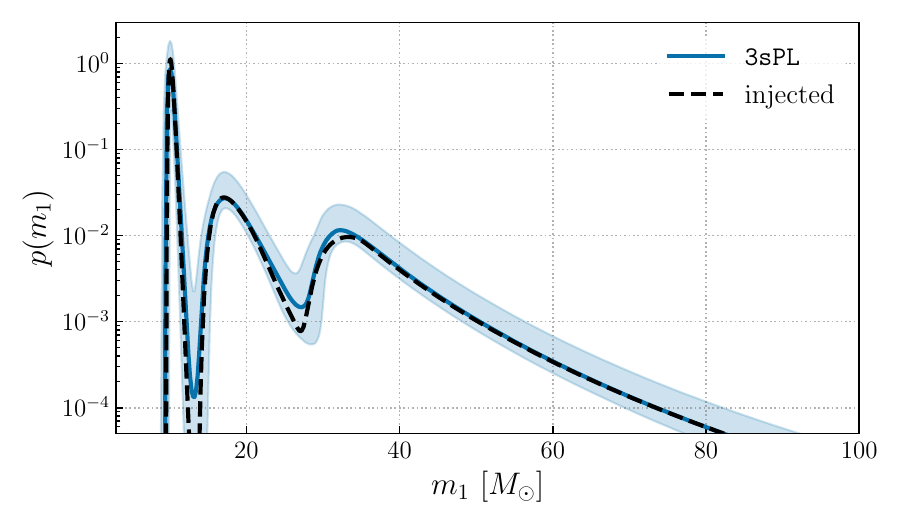}
        \caption{}
        \label{fig:O5_LCDM_PPD}
    \end{subfigure}
    \caption[O5 - \LCDM]{
    \textbf{(a)} O5 forecasts for constraints on parameters of the \texttt{FlatLCDM} model and \textbf{(b)} primary mass \ac{PPD} (with 90\% \ac{CI}), for a mock population generated from a \camelthrees distribution and recovered with a \camelthrees in blue, solid when $\Omega_m$ is inferred, dashed when fixed. 
    Solid and dashed black lines, on the \textbf{(a)} and \textbf{(b)} subfigures respectively, indicate the value injected in simulations. 2D contours indicate 90\% credible area.}
    \label{fig:O5_LCDM}
\end{figure}

\subsubsection{\texorpdfstring{\LCDM}{LCDM}}
\label{sec:res:O5:LCDM}

Fig.~\ref{fig:O5_LCDM} summarises the results for \LCDM. We find a constraint on the Hubble constant of $H_0 = 75.6^{+11.4}_{-10.7} ~\rm km \,s^{-1} \,Mpc^{-1}$ (solid lines in the figure), corresponding to a precision of 15\%, a substantial improvement compared to similar studies in the literature. Ref.~\cite{Mancarella:2021ecn} forecasts a $\sim 20\%$ precision by analysing $\sim 5000$ simulated events with $\rm SNR > 8$ drawn from a broken power-law population; Refs.~\cite{Leyde:2022orh} and \cite{Borghi:2023opd} find $\sim 24\%$ and $\sim 32\%$ from $\sim 500$ simulated events with $\rm SNR > 12$ (same as ours) and 100 events with $\rm SNR > 25$, respectively, using a more realistic power-law + Gaussian population. While Ref.~\cite{Mancarella:2021ecn} uses a simulation setup similar to ours with full individual events \ac{PE}, Refs.~\cite{Leyde:2022orh} and \cite{Borghi:2023opd} employ simpler methods which only approximate full correlations between events' parameters. Beyond these simulated dataset differences, our more optimistic constraints primarily arise from the inclusion of additional, sharper features in the mass distribution, similarly to the results on real data. This scenario is so promising that the presence of exceptionally sharp features would enable $\sim 3\%$ measurements from the population alone. This possibility was first suggested using the pair-instability mass gap~\cite{Farr:2019twy}, and more recently demonstrated with a sharp peak at $10~\rm M_{\odot}$ that remains fully compatible with current data~\cite{Bertheas:2025mzd}.

The lower-left panel of Fig.~\ref{fig:O5_LCDM} also highlights the more pronounced $H_0$–$\Omega_m$ correlation. When $\Omega_m$ is fixed, the constraint on $H_0$ improves to about $6\%$ (dashed blue line), illustrating how much of the uncertainty is driven by this degeneracy.
The broader redshift range covered by the O5 catalog also allows us to place constraints on $\Omega_m$, yielding a measurement of $\Omega_m = 0.29^{+0.23}_{-0.15}$, corresponding to a precision of $60\%$. This represents a significant improvement compared to the essentially uninformative posteriors obtained from \ac{GWTC} data.
We emphasize, however, that breaking the $H_0$–$\Omega_m$ degeneracy requires substantial information from the data, and further improvements in the measurement of $H_0$ are therefore largely limited by how well $\Omega_m$ can be constrained (see, e.g.,~\cite{Bertheas:2025mzd}).
Finally, note that small shifts of the $H_0$ posterior around the injected value is expected from Poisson noise in that specific population realisation.

\begin{figure}[!b]
    \centering
    \includegraphics[width=0.6\linewidth]{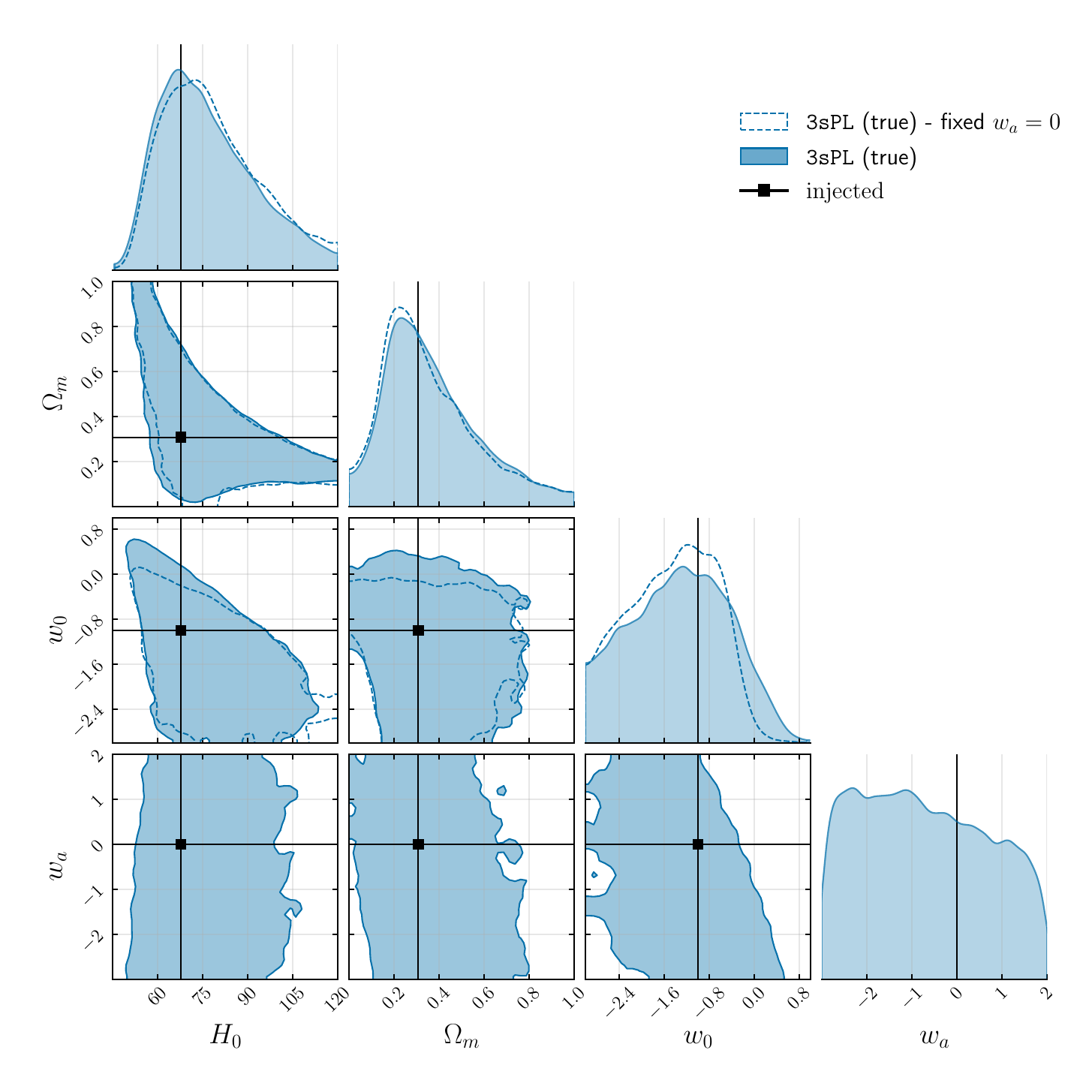}
    \caption[O5 - \wowaCDM - Inferred: $H_0, \Omega_m, w_0, w_a$.]{O5 forecasts for constraints on parameters of \texttt{w0waCDM} model, for a mock catalog simulated and analysed with a \camelthrees population. Solid (resp.~dashed) lines denote results of analyses with inferred (resp.~fixed) $w_a$. Solid black lines indicates the value injected in simulations. 2D contours indicate the 90\% credible areas.}
    \label{fig:O5_w0waCDM}
\end{figure}

\subsubsection{Dark energy}
\label{sec:res:O5:DE}

For dynamical \ac{DE}, the results in Fig.~\ref{fig:O5_w0waCDM} show that the constraints are only marginally improved compared to the current \ac{GWTC} accuracy.
In particular, while the posterior on $w_0$ weakly peaks at its injected value of $-1$, posterior on $w_a$ remains uninformative in O5. 
However, compared to \ac{GWTC} data, we observe an increased correlation between $H_0$ and $w_0$, reflecting the exclusion of a portion of the parameter space. This weakens the constraints on the Hubble constant compared to the \LCDM case (compare to Fig.~\ref{fig:O5_LCDM}). The measurement of $\Omega_m$ remains consistent with the \LCDM case as the analysis exhibits no clear correlation between $\Omega_m$ and \ac{DE} parameters. Restricting the analysis to the case of a non-evolving \ac{DE} equation of state (i.e. fixing $w_a = 0$), we find a minor improvement in the $w_0$ posterior, although still mostly uninformative. The persistent absence of significant constraints on \ac{DE} parameters, even at O5 sensitivity, may be attributed to several factors. These include, for example, the lack of pronounced structure at the high-mass end of the population (as extensively discussed in Sec.~\ref{sec:res:gwtc4}), as well as the choice of priors adopted in this work. While these priors are designed to explore only physically motivated values of the \ac{DE} equation of state~\cite{DESICollaboration2025ddr}, they may restrict the parameter space to regions where deviations from \LCDM\ remain undetectable with current-generation detectors, contrary to those adopted for modified \ac{GW} propagation parameters. Further studies are needed to determine the conditions required to constrain \ac{DE} with \acp{GW}.

\begin{figure}[!b]
    \centering
    \begin{subfigure}[c]{0.49\textwidth}
        \centering
        \includegraphics[width=\linewidth]{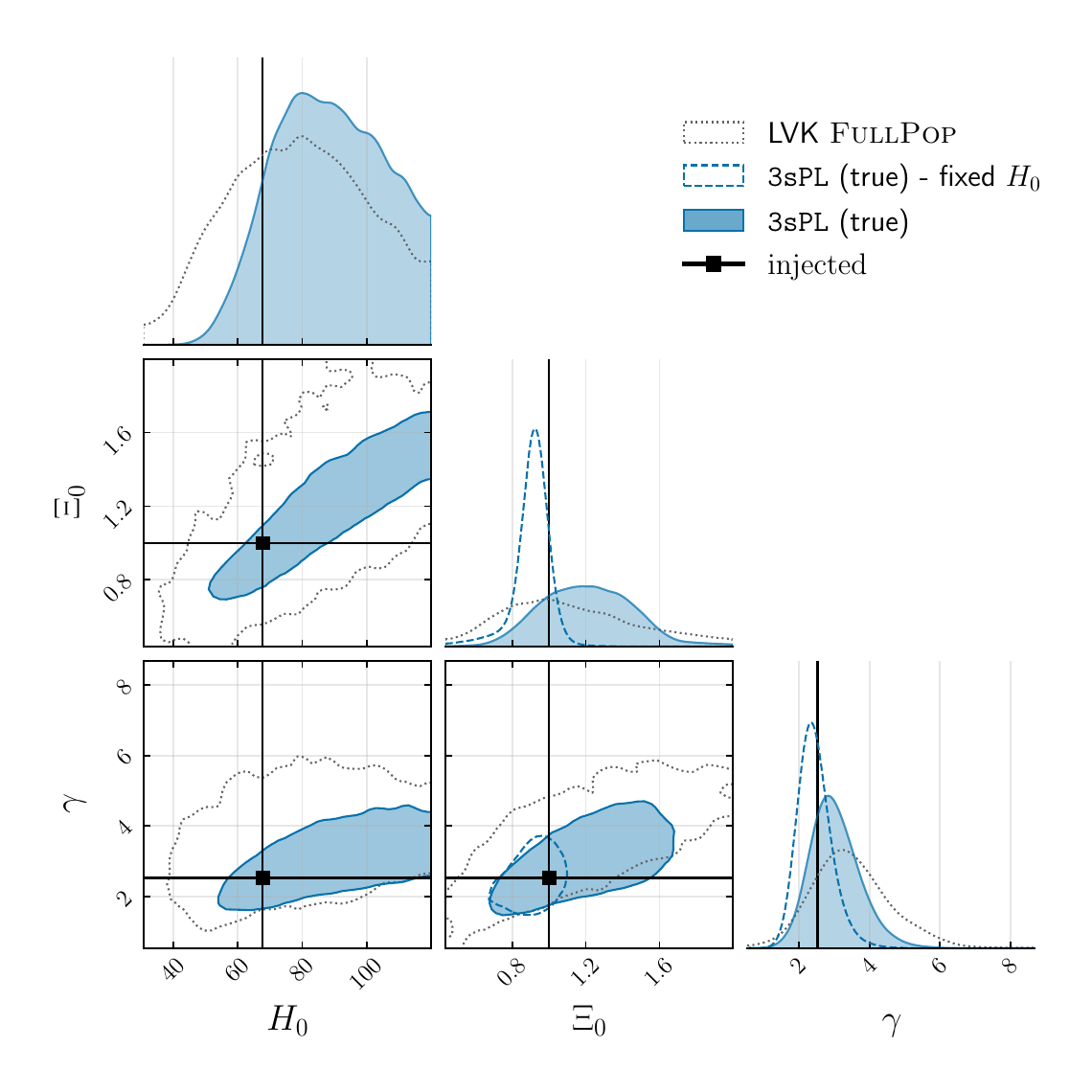}
        \caption{}
        \label{fig:O5_Xi0}
    \end{subfigure}
    \begin{subfigure}[c]{0.49\textwidth}
        \centering
        \includegraphics[width=\linewidth]{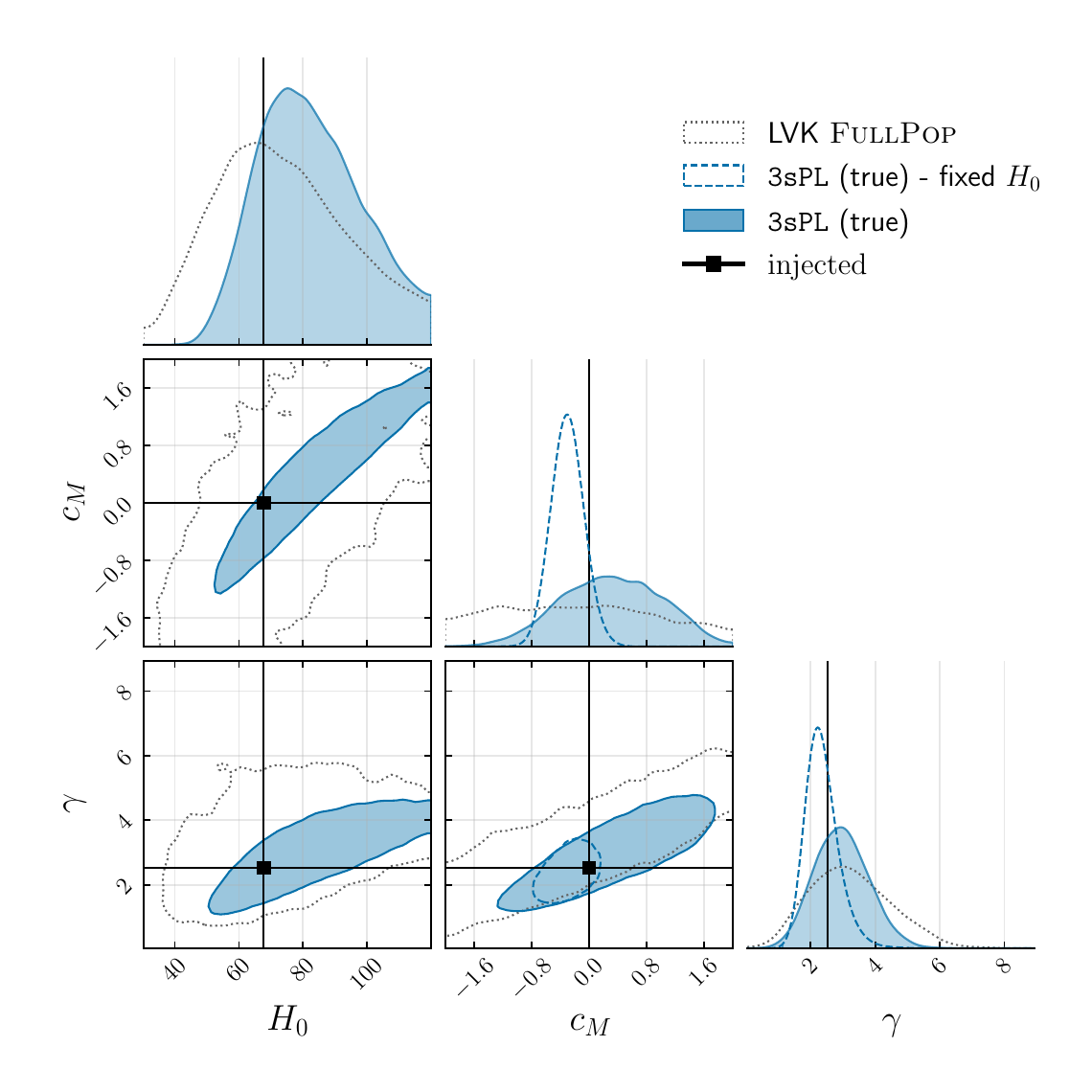}
        \caption{}
        \label{fig:O5_cM}
    \end{subfigure}
    \caption[O5 - MG]{O5 forecasts for constraints and correlations on parameters of modified \ac{GW} propagation models \texttt{Xi0} \textbf{(a)} and \texttt{cM} \textbf{(b)}, for a mock catalog simulated and analysed with a \camelthrees population. Solid (dashed) lines posteriors denote results of analyses with inferred (fixed) $H_0$. Solid black lines indicates the value injected in simulations. 2D contours indicate the 90\% credible areas.}
    \label{fig:O5_beyond-GR}
\end{figure}

\subsubsection{Modified GW propagation}
\label{sec:res:O5:MG}

Finally, we observe a significant improvement in modified \ac{GW} propagation models when going from \ac{GWTC} data to O5 prospects.
As can be seen in Fig.~\ref{fig:O5_beyond-GR}, we obtain $\Xi_0 = 1.20^{+0.25}_{-0.24}$ and $c_M = 0.3^{+0.7}_{-0.7}$, corresponding to improvements of a factor $\sim 6$ and $\sim 3$, respectively, relative to the \ac{GWTC} results (where jointly inferred with $H_0$).
In the \texttt{Xi0} model case, $n$ is still unconstrained.
This is expected from the \ac{GR} universe we simulated with $\Xi_0 = 1$, which is completely degenerate over values of $n$.
As expected, the increased detector sensitivity strengthens the correlations between modified \ac{GW} propagation parameters and $H_0$, highlighting the well-known degeneracy between them~\cite{Mancarella:2021ecn, Leyde:2022orh, Tagliazucchi:2025ofb}. Fixing $H_0$ therefore leads to substantially tighter constraints, $\Xi_0 = 0.92^{+0.06}_{-0.07}$ and $c_M = -0.3^{+0.2}_{-0.2}$, improving by factors of $\sim2$ and $\sim3$ relative to the marginalised case.
In terms of measurement precision, this corresponds to about $20\%$ ($7\%$) on $\Xi_0$ when inferring (fixing) $H_0$. 
These results can be compared with several forecasts in the literature.
Ref.~\cite{Leyde:2022orh} predicts a $29\%$ precision after marginalising over $H_0$ and $\Omega_m$, analysing $\sim 500$ events with $\rm SNR > 12$ drawn from a power-law + Gaussian population.
Ref.~\cite{Mancarella:2021ecn} finds $20\%$ ($10\%$) precision when marginalising over (fixing) both $H_0$ and $\Omega_m$, using $\sim 5000$ events with $\rm SNR > 8$ drawn from a broken power-law population.
Finally, Ref.~\cite{Tagliazucchi:2025ofb} forecasts $8\%$ ($41\%$) precision when marginalising over $H_0$ with fixed $\Omega_m$, analysing $\sim 400$ high-SNR events with $\rm SNR > 20$ from a power-law + Gaussian population, combined with a mock galaxy catalog providing spectroscopic (photometric) redshifts.

In the spectroscopic case of~\cite{Tagliazucchi:2025ofb}, the galaxy information enables a sub-percent determination of $H_0$, effectively equivalent to fixing it; our forecasts with fixed $H_0$ are therefore comparable to that scenario.
Conversely, the photometric-redshift case yields significantly weaker $H_0$ constraints ($27\%$ precision), and the strong correlation with $\Xi_0$ correspondingly degrades the measurement of the latter.
The more optimistic constraints obtained in our analysis are mainly driven by the \ac{BBH} population model we adopt, which contains additional and sharper structure in the mass distribution, as well as by the larger number of events in the dataset. In particular, the pronounced low-mass structure improves the determination of $H_0$ (to $20\%$ precision), even without incorporating any galaxy catalog information, which in turn tightens the constraints on $\Xi_0$.

\section{Conclusion}
\label{sec:conclusion}
In this work, we provide the best constraints to date on cosmological parameters employing the spectral sirens method with \acp{BBH}-only events from the most recent \ac{GWTC} data released by the \ac{LVK} collaboration.
We do so by introducing new parametric primary mass models built as linear combinations of 3 and 4 truncated, tapered power-law components (\camelthrees and \camelfours), which are able to resolve sharper features in the primary mass distribution of \ac{GWTC} events compared to other parametric models in the literature~\cite{LIGOScientific:2025pvj, LIGOScientific:2025jau}.
Specifically, the reconstructed distributions highlight the presence of a rich structure at low mass, with a sharp peak at $10 ~\rm M_\odot$ followed by a second overdensity at $\sim 20 ~\rm M_\odot$ and a possible gap in between~\cite{Toubiana:2023egi, Gennari:2025nho, Tiwari:2025lit, Willcox:2025poh}. Besides the $\sim 35 ~\rm M_\odot$ peak, further structure at high mass is captured by the \camelfours model, in agreement with previous studies~\cite{LIGOScientific:2025pvj, MaganaHernandez:2024qkz, Pierra:2026ffj}.
As for previous catalogs~\cite{Gennari:2025nho, Bertheas:2025mzd}, we show that the \camelthrees model is mildly favoured 
by the \ac{GWTC} data over power-law + Gaussians models similar to those used by \ac{LVK}~\cite{LIGOScientific:2021aug, LIGOScientific:2025jau} (see Tabs.~\ref{tab:gwtc4_constraints:GR} and \ref{tab:gwtc4_constraints:MG}). Notice, however, that correcting the \acp{BF} for the effective prior volume explored during inference tends to marginally lower the evidence in favor of \camelthrees (more details can be found in App.~\ref{app:numerical_stability}). As a result, despite the larger number of parameters, our power-law based models provide a statistically equivalent description of the \ac{BBH} population. In addition, some of these parameters, such as the maximum mass for the power-law components, remain unconstrained by the data and therefore contribute negligibly to the Bayesian evidence.

Adopting a \LCDM cosmological model, we constrain the Hubble constant to $H_0 = 53.3^{+14.0}_{-10.8} ~\rm km \,s^{-1} \,Mpc^{-1}$, corresponding to a precision of 23\%. 
We additionally find that enforcing a shared minimum mass for the primary and secondary distributions prevents the reconstruction of the underdensity following the $10 ~\rm M_\odot$ peak within the parametric models considered, instead favouring support at lower masses over a sharp cutoff. Nonetheless, these significant differences in the reconstructed distribution do not affect the $H_0$ measurements, whose posteriors peak at lower values than those associated with the Hubble-tension region. This shift of the $H_0$ posterior towards lower values has already been found in independent works~\cite{Mould:2026sww, Tagliazucchi:2026gxn}, despite different mass models being adopted there.
Our results highlight the importance of population modelling for spectral sirens, which can significantly improve cosmological inference. However, fully parametric modelling is known to be susceptible to systematic errors~\cite{Romero-Shaw:2022ctb, Pierra:2023deu, Gennari:2025nho, Agarwal:2024hld}, raising the question of introducing significant biases in future runs with improved sensitivity. These issues could potentially be mitigated by using more agnostic semi- or non-parametric methods (e.g.~\cite{Tagliazucchi:2026gxn}). A detailed study of this subject will be presented in a separate publication.

We investigate different cosmological models, and report no significant constraints on the \ac{DE} equation of state parameters, $(w_0,w_a)$.
Regarding modified \ac{GW} propagation, we find comparable level of constraints on $\Xi_0$ and $c_M$ parameters using \acp{BBH} events only --- namely $\Xi_0 = 1.4^{+0.9}_{-0.5}$ and $c_M = 0.7^{+1.7}_{-1.7}$ with the \camelfours mass model --- to the ones reported by the \ac{LVK} collaboration using all \ac{CBC} events combined with galaxy catalogs. As \ac{GW} propagation parameters are more sensitive to information from high redshifts, our results suggest that mass models capable of capturing the structure of the high-mass population spectrum can improve these measurements.

Finally, we simulate a mock catalog based on the \camelthrees mass model in order to provide forecasts on the measurement of cosmological parameters at O5 sensitivity.
We find a $\sim 15\%$ precision on $H_0$ for a \texttt{FlatLCDM} cosmology, which improves to $\sim 6\%$ when fixing $\Omega_m$ (mimicking the combination of \acp{GW} with \ac{EM} probes more sensitive to matter density parameters). 
While our results on dynamical \ac{DE} parameters suggest that constraints on such models will remain beyond the reach of the spectral sirens method with the current generation of \ac{GW} detectors, we nonetheless forecast an improvement in the constraints on modified \ac{GW} propagation parameters by a factor 3 to 6 compared to current data.

A crucial assumption of our work is the stationarity of the \ac{BBH} mass distribution. 
So far, a wide range of methods has found no significant evidence for intrinsic redshift evolution~\cite{LIGOScientific:2025pvj, Sadiq:2021fin, Ray:2023upk, Heinzel:2024hva, Mali:2024wpq, Gennari:2025nho, Lalleman:2025xcs, Sadiq:2025aog} (although see~\cite{Rinaldi:2023bbd}), even though current observations are not yet sufficiently informative to draw robust conclusions. 
However, unmodelled mass–redshift correlations are known to bias cosmological measurements~\cite{Pierra:2023deu, Agarwal:2024hld, Mali:2024wpq}. 
Although we include event-rate evolution, unmodelled mass–redshift correlations expected at the higher redshifts probed in our simulations are likely to primarily affect the results in Sec.~\ref{sec:res:O5}. 
Similarly to several analyses in the literature about future LVK observing runs and even 3G detectors~\cite{Califano:2025qbx, Tagliazucchi:2026gxn, Borghi:2023opd, Leyde:2022orh}, our forecasts should therefore be regarded as optimistic in the absence of a comprehensive understanding of the interplay between cosmological and mass-evolution parameters.

Although the present work focuses on extracting information on cosmological parameters from \ac{GW} data alone, our analysis framework on \ac{GWTC} data could further include \ac{EM} information from galaxy catalogs, as routinely done in related works~\cite{Gair:2022zsa, Gray:2023wgj, LIGOScientific:2025jau}. 
We expect this to only marginally improve on the results presented here, as current dark sirens analyses are mostly driven by the population~\cite{LIGOScientific:2025jau}.
We also note that our constraints are comparable to those obtained from the one bright siren event observed so far~\cite{LIGOScientific:2017adf, LIGOScientific:2018hze, Palmese:2023beh, LIGOScientific:2025jau}. 
Similar considerations apply to the O5 forecasts, where our constraints could be further enhanced by combining \acp{GW} with galaxy catalogs. As demonstrated in previous studies~\cite{Borghi:2023opd, Tagliazucchi:2025ofb}, complete galaxy catalogs alone can push cosmological measurements to percent level by O5, establishing \acp{GW} as a compelling scientific case for independently resolving the Hubble tension. On the other hand, it has recently been shown that the same conclusion can be obtained from sharp features in the BH distribution alone~\cite{Bertheas:2025mzd}, forecasting optimistic population models that are yet fully compatible with current data.
To quantify the relative contribution of the population and the galaxy catalogs that we will achieve in a few years, more realistic simulations with incomplete galaxy catalogs will be needed.

\acknowledgments

The authors thank G.~Pierra, S.~Mastrogiovanni, A.~Papadopoulos, K.~Leyde and M.~Mould for insightful discussions, as well as the anonymous referee for useful comments.

T.B., V.G. and N.T. acknowledge support form the French space agency CNES in the framework of LISA. 
T.B. acknowledges support from a CDSN PhD grant from ENS-PSL.
This project has received financial support from the Agence Nationale de la Recherche (ANR) through the MRSEI project ANR-24-MRS1-0009-01.
The authors are grateful for computational resources provided by the IN2P3 computing centre (CC-IN2P3) in Lyon (Villeurbanne).
This research made use of data, software and/or web tools obtained from the Gravitational Wave Open Science Center~\cite{LIGOScientific:2019lzm, KAGRA:2023pio}, a service of the LIGO Scientific Collaboration, the KAGRA Collaboration and the Virgo Collaboration.
This material is based upon work supported by NSF's LIGO Laboratory which is a major facility fully funded by the National Science Foundation.

\newpage
\appendix
\section{Models and priors}
\label{app:models_priors}

\subsection{Population models}
\label{app:models_priors:pop}

\subsubsection{Basic probability distributions}

Most of the mass models are built as combinations of the following basic probability distributions:

\paragraph{Truncated power-law:}  

\begin{equation}
    \label{eq:trPL}
     \mathcal{P}(x | x_{\rm min}, x_{\rm max}, \alpha) =
    \begin{cases}
        \frac{1}{N} x^{\alpha} & \qq{\rm if} x_{\rm min} < x < x_{\rm max} \\
        0 & \qq{\rm otherwise}
    \end{cases}
\end{equation}
The normalization factor is given by:
\begin{equation}
\label{eq:PL_norm}
    N =
    \begin{cases}
        \frac{1}{\alpha + 1} \left( x_{\rm max}^{\alpha + 1} - x_{\rm min}^{\alpha + 1} \right) & \qq{\rm if} \alpha \neq -1 \\
        \ln{ \frac{x_{\rm max}}{x_{\rm min}} } & \qq{\rm otherwise}
    \end{cases}
\end{equation}

\paragraph{Truncated Gaussian:}

\begin{equation}
\label{eq:trGauss}
    \mathcal{G}_{[x_{\rm min}, x_{\rm max}]}(x | \mu, \sigma) =
    \begin{cases}
        \frac{1}{N} \frac{1}{ \sigma \sqrt{2 \pi} } e^{- \frac{ (x - \mu)^2 }{2 \sigma^2}} & \qq{\rm if} x_{\rm min} < x < x_{\rm max} \\
        0 & \qq{\rm otherwise}
    \end{cases}
\end{equation}
The normalization factor is expressed through the $\erf$ function:
\begin{equation}
\label{eq:Gauss_norm}
N = \frac{1}{2} \left[ \erf \left( \tfrac{x_{\rm max} - \mu}{\sigma \sqrt{2}} \right) - \erf \left( \tfrac{x_{\rm min} - \mu}{\sigma \sqrt{2}} \right) \right]
\end{equation}

\paragraph{High-pass filter:} 
\begin{equation}
\label{eq:smoothing}
    \mathcal{S}(x | x_{\rm min}, \delta) = 
    \begin{cases}
        0, & \qq{\rm if} x < x_{\rm min} \\
        \left( f(x - x_{\rm min}, \delta) + 1 \right)^{-1}, & \qq{\rm if} x_{\rm min} \leq x < x_{\rm min}+\delta \\
        1, & \qq{\rm if} x_{\rm min} + \delta \leq x
    \end{cases}
\end{equation}
with
\begin{equation}
\label{eq:smoothing_sig}
    f(x', \delta) = \exp \left(\frac{\delta}{x'} + \frac{\delta}{x' - \delta}\right).
\end{equation}
When we apply this window, the distributions are numerically re-normalized. Note that at $x = x_{\rm min} + 0.5 \delta_m$ the filter is 0.5.

Note that this tapering function is mainly applied to power-law components, so we define the tapered power-law $\mathcal{SP}$ distribution as follows.
\begin{equation}
\label{eq:sPL}
    \mathcal{SP}(x | x_{\rm min}, x_{\rm max}, \alpha, \delta) \propto \mathcal{S}(x | x_{\rm min}, \delta) \times \mathcal{P} (x | x_{\rm min}, x_{\rm max}, \alpha)
\end{equation}
which is properly renormalized after applying the tapering to the power-law.

\subsubsection{Primary mass models}

\paragraph{\texttt{smooth Power Law + 2 Gaussians (sPL2G)}} This model consists of one power-law component to which we add two Gaussian, the whole distribution being tapered at its lower end:
\begin{equation}
\label{eq:sPL2G}
\begin{aligned}
    \pi_{\sPLGG}(m_{1,s} | \Lambda_{\sPLGG}) = & \, \mathcal{S}(m_{1,s}| m_{\rm min}, \delta_m) \Big[ (1 - \lambda_{\rm g}) \mathcal{P}(m_{1,s}| m_{\rm min}, m_{\rm max}, - \alpha) \\
    & + \lambda_{\rm g} \lambda_{{\rm g}, a} \mathcal{G}_{[m_{\rm min}, \, \mu_{{\rm g}, a} + 5 \sigma_{{\rm g}, a}]} (m_{1,s} | \mu_{{\rm g}, a}, \sigma_{{\rm g}, a}) \\
    & + \lambda_{\rm g} (1 - \lambda_{{\rm g}, a}) \mathcal{G}_{[m_{\rm min}, \, \mu_{{\rm g}, b} + 5 \sigma_{{\rm g}, b}]} (m_{1,s} | \mu_{{\rm g}, b}, \sigma_{{\rm g}, b}) \Big] \\
\end{aligned}    
\end{equation}
with parameters
\begin{equation}
\label{eq:pars_sPL2G}
    \Lambda_{\sPLGG} = (\alpha, m_{\rm min}, m_{\rm max}, \mu_{{\rm g}, a}, \sigma_{{\rm g}, a}, \mu_{{\rm g}, b}, \sigma_{{\rm g}, b}, \delta_m, \lambda_{\rm g}, \lambda_{{\rm g}, a})
\end{equation}
detailed in Tab.~\ref{tab:priors_sPL2G}.

\begin{table}[t]
    \centering
    \begin{tabular}{@{\extracolsep\fill}lcc}
        \toprule%
        \multirow{2}{*}{\textbf{\sPLGG parameters}} & \multicolumn{2}{@{}c@{}}{\textbf{Priors}} \\ \cmidrule{2-3}
         & \textbf{GWTC-4} & \textbf{O5 mock data} \\
        \midrule
        $\alpha$ & (1.5, 8) & (0, 200) or (0, 6) \\
        $m_{\rm min} ~\rm[M_\odot]$ & (2, 10) & (1, 40) or (1, 20) \\
        $m_{\rm max} ~\rm[M_\odot]$ & (50, 200) & (50, 200) \\
        $\delta_m ~\rm[M_\odot]$ & ($10^{-3}$, 10) & (0, 30) \\
        $\mu_{{\rm g}, a} ~\rm[M_\odot]$ & (5, 20) & (1, 60) or (5, 15) \\
        $\sigma_{{\rm g}, a} ~\rm[M_\odot]$ & ($10^{-3}$, 5) & ($10^{-3}$, 50) or ($10^{-3}$, 2) \\
        $\mu_{{\rm g}, b} ~\rm[M_\odot]$ & (20, 100) & (1, 60) \\
        $\sigma_{{\rm g}, b} ~\rm[M_\odot]$ & ($10^{-3}$, 10) & ($10^{-3}$, 50) or ($10^{-3}$, 10) \\
        $\lambda_{\rm g}$ & (0, 1) & (0, 1) \\
        $\lambda_{{\rm g}, a}$ & (0, 1) & (0, 1) \\
        \bottomrule
    \end{tabular}
    \caption[Priors - \sPLGG]{Prior probability distributions assumed on parameters of the \sPLGG primary mass distribution model, for both the analysis of GWTC-4 data and O5 simulated data. All priors are uniform in the given ranges.}
    \label{tab:priors_sPL2G}
\end{table}

As explained above, the distribution is renormalised when applying the smoothing function $\mathcal{S}$, hence $\pi_{\sPLGG}$ being normalized.

\paragraph{\texttt{Power Law mixture models (NsPL} with \texttt{N = 3, 4)}} This model consists of a combination of several power-law components tapered at their lower end, with parameters listed in Tab.: 
\begin{equation}
\label{eq:NsPL}
\begin{aligned}
    \pi_{\camels}(m_{1,s} | \Lambda_{\camels}) = & \, \sum_{x \in {\{\rm latin\}}_{\texttt{N}-1}} \lambda_{x} \,\mathcal{SP}(m_{1,s}| m_{{\rm 1, min}, x}, m_{{\rm 1, max}, x}, - \alpha_{x}, \delta_{m_1, x}) \\
    & + \left( 1 - \sum_{x \in {\{\rm latin\}}_{\texttt{N}-1}} \lambda_{x} \right) \,\mathcal{SP}(m_{1,s}| m_{{\rm 1, min}, a_\texttt{N}}, m_{{\rm 1, max}, a_\texttt{N}}, - \alpha_{X}, \delta_{m_1, a_\texttt{N}}) \\
\end{aligned}    
\end{equation}
with parameters
\begin{equation}
\label{eq:pars_NsPL}
\begin{split}
    \Lambda_{\camels} = ~& (\alpha_a, \dots, \alpha_{{a_\texttt{N}}}, ~m_{{\rm 1, min}, a}, \dots, m_{{\rm 1, min}, {a_\texttt{N}}}, ~m_{{\rm 1, max}, a}, \dots, m_{{\rm 1, max}, {a_\texttt{N}}}, \\ & \delta_{m_1, a}, \dots, \delta_{m_1, {a_\texttt{N}}}, ~\lambda_1, \dots, \lambda_{a_{\texttt{N}-1}})
\end{split}
\end{equation}
detailed in Tab.~\ref{tab:priors_NsPL}. The notation ${\{\rm latin\}}_{\texttt{N}}$ refers to the set of $\texttt{N}$ first letters in the latin alphabet, while $a_\texttt{N}$ denotes the $\texttt{N}^{\rm th}$ letter.

\begin{table}[t]
    \centering
    \begin{tabular}{@{\extracolsep\fill}lccc}
        \toprule%
        \multirow{2}{*}{\textbf{\camels parameters}} & \multicolumn{2}{@{}c@{}}{\textbf{Priors}} & \textbf{Simulations} \\ \cmidrule{2-3} \cmidrule{4-4}
         & \textbf{GWTC-4} & \textbf{O5 mock data} & \textbf{injected $\Lambda_{\texttt{3sPL}}$} \\
        \midrule
        $\alpha_a$ & (0, 200) & (0, 200) & 59.9 \\
        $m_{{\rm min}, a} ~\rm[M_\odot]$ & (5, 15) & (8, 11) & 9.3 \\
        $m_{{\rm max}, a} ~\rm[M_\odot]$ & (15, 30) & (20, 25) & 22.7 \\
        $\delta_{m_1, a} ~\rm[M_\odot]$ & ($10^{-3}$, 20) & ($10^{-3}$, 10) & 3.08 \\
        $\alpha_b$ & (0, 30) & (0, 30) & 10.8 \\
        $m_{{\rm min}, b} ~\rm[M_\odot]$ & (10, 20) & (10, 20) & 12.9 \\
        $m_{{\rm max}, b} ~\rm[M_\odot]$ & (20, 100) & (67.5, 72.5) & 69.9 \\
        $\delta_{m_1, b} ~\rm[M_\odot]$ & ($10^{-3}$, 50) & ($10^{-3}$, 50) & 10.8 \\
        $\alpha_c$ & (0, 30) & (0, 30) & 6.1 \\
        $m_{{\rm min}, c} ~\rm[M_\odot]$ & (15, 40) & (20, 30) & 25.5 \\
        $m_{{\rm max}, c} ~\rm[M_\odot]$ & (40, 150) & (101.5, 106.5) & 104.1 \\
        $\delta_{m_1, c} ~\rm[M_\odot]$ & ($10^{-3}$, 50) & ($10^{-3}$, 50) & 11.9 \\
        $\alpha_d$ & (0, 30) & (0, 200) & - \\
        $m_{{\rm min}, d} ~\rm[M_\odot]$ & (35, 80) & (5, 50) & - \\
        $m_{{\rm max}, d} ~\rm[M_\odot]$ & (80, 150) & (20, 120) & - \\
        $\delta_{m_1, d} ~\rm[M_\odot]$ & ($10^{-3}$, 50) & ($10^{-3}$, 50) & - \\
        $\lambda_a$ & (0, 1) & (0, 1) & 0.741 \\
        $\lambda_b$ & (0, 1) & (0, 1) & 0.147 \\
        $\lambda_c$ & (0, 1) & (0, 1) & - \\
        \bottomrule
    \end{tabular}
    \caption[Priors - \camels]{Prior probability distributions assumed on parameters of the \camelthrees and \camelfours primary mass distribution models, for both the analysis of GWTC-4 data and O5 simulated data. All priors are uniform in the given ranges. The last column shows approximate values of the injected parameters in the simulations at O5 sensitivity.}
    \label{tab:priors_NsPL}
\end{table}

Note that since $\mathcal{SP}$ is already normalized, $\pi_{\camels}$ is automatically normalized.

\subsubsection{Mass ratio model}
\label{app:models:q}

\begin{table}[!b]
    \centering
    \begin{tabular}{@{\extracolsep\fill}lccc}
        \toprule%
        \multirow{2}{*}{\makecell[l]{\textbf{Mass ratio}\\\textbf{parameters}}} & \multicolumn{2}{@{}c@{}}{\textbf{Priors}} & \textbf{Simulations} \\ \cmidrule{2-3} \cmidrule{4-4}
         & \textbf{GWTC-4} & \textbf{O5 mock data} & \textbf{injected $\Lambda_q$} \\
        \midrule
        $\mu_q$ & (0.5, 1) & (0.5, 1) & 0.85 \\
        $\sigma_q$ & (0.01, 0.5) & (0.01, 0.5) & 0.21 \\
        \bottomrule
    \end{tabular}
    \caption[Priors - mass ratio]{Prior probability distributions assumed on parameters of the mass ratio distribution models, for both the analysis of GWTC-4 data and O5 simulated data. All priors are uniform in the given ranges. The last column shows approximate values of the injected parameters in the simulations at O5 sensitivity.}
    \label{tab:priors_q}
\end{table}

The mass ratio $q = m_{2,s} / m_{1,s} \in [0, 1]$ distribution is parametrised as truncated Gaussian, with parameters $\Lambda_{q} = (\mu_q, \sigma_q)$ listed in Tab.~\ref{tab:priors_q}:
\begin{equation}
\label{eq:model_q}
    \pi_q(q | \Lambda_{q}) = \mathcal{G}_{[0, 1]} (q | \mu_q, \sigma_q) \,.
\end{equation}

\subsubsection{Secondary mass models}
\label{app:models:m2}

Sec.~\ref{sec:res:gwtc4:m2param} explores an alternative approach in which the secondary-mass distribution $m_{2,s}$ is parametrised as a truncated power law conditioned on the primary mass $m_{1,s}$ (to account for the usual $m_{2,s} < m_{1,s}$ ordering convention in \ac{GW} data analysis) and tapered at the low-mass end, with parameters $\Lambda_{m_2} = (\beta, m_{\rm 2, min}, \delta_{m_2})$:
\begin{equation}
\label{eq:model_m2}
    \pi(m_{2,s} | m_{1,s}) = \mathcal{SP}(m_{2,s} | m_{\rm 2, min}, m_{1,s}, \beta, \delta_{m_2}) \,.
\end{equation}
In the main text, this model is only considered in combination with \camelthrees for the primary mass. We study two cases: $(m_{\rm 2, min}, \delta_{m_2}) = (m_{{\rm 1, min}, a}, \delta_{m_1, a})$ and $(m_{\rm 2, min}, \delta_{m_2}) \neq (m_{{\rm 1, min}, a}, \delta_{m_1, a})$, corresponding to a scenario in which the low-mass end of the spectrum is jointly informed by both primary and secondary \ac{BBH} components, and one in which the low-mass ends of the primary and secondary mass distributions are inferred independently.

\subsubsection{Rate model}

The rate evolution is parametrised with a function matching the Madau-Dickinson star formation rate, with parameters $\Lambda_{\texttt{MD}} = (\gamma, \kappa, z_p)$ listed in Tab.~\ref{tab:priors_MD}:
\begin{equation}
\label{eq:madau_diskinson_SFR}
    \psi_{\texttt{MD}}(z | \Lambda_{\texttt{MD}}) = \left[ 1 + \left( \frac{1}{1 + z_p} \right)^{\gamma + k} \right] \frac{ (1 + z)^\gamma }{ 1 + \left( \frac{1 + z}{1 + z_p} \right)^{\gamma + k} }.
\end{equation}

\begin{table}[!h]
    \centering
    \begin{tabular}{@{\extracolsep\fill}lccc}
        \toprule%
        \multirow{2}{*}{\textbf{\texttt{MD} parameters}} & \multicolumn{2}{@{}c@{}}{\textbf{Priors}} & \textbf{Simulations} \\ \cmidrule{2-3} \cmidrule{4-4}
         & \textbf{GWTC-4} & \textbf{O5 mock data} & \textbf{injected $\Lambda_q$} \\
        \midrule
        $\gamma$ & (-5, 10) & (-5, 10) & 2.53 \\
        $\kappa$ & (-5, 10) & (-5, 10) & 3.05 \\
        $z_p$ & (0, 4) & (0, 4) & 1.49 \\
        \bottomrule
    \end{tabular}
    \caption[Priors - \texttt{MD}]{Prior probability distributions assumed on parameters of the Madau-Dickinson rate evolution model \texttt{MD}, for both the analysis of GWTC-4 data and O5 simulated data. All priors are uniform in the given ranges. The last column shows approximate values of the injected parameters in the simulations at O5 sensitivity.}
    \label{tab:priors_MD}
\end{table}

\subsection{Cosmological models}
\label{app:models_priors:cosmo}

This appendix mainly contains Tab.~\ref{tab:priors_cosmo} summarising the priors imposed on cosmological parameters throughout the present paper, for cosmological models described in Sec.~\ref{sec:methods:cosmo}. 
For the \texttt{w0waCDM} model, one may choose to restrict the \ac{DE} parameter space to the physical region $w_0 + w_a < 0$, which ensures an early matter-dominated era~\cite{DESICollaboration2025ddr}. 
We have also performed analyses imposing this condition as a prior on the \ac{DE} equation-of-state parameters, finding no impact on our conclusions. In the main text, we therefore present only results obtained with agnostic priors, as such constraints can lead to misleading 1D posteriors, where apparent structure may be driven by the prior rather than by the data.

\begin{table}[!h]
    \centering
    \begin{tabular}{@{\extracolsep\fill}lcc}
        \toprule%
        \multirow{2}{*}{\makecell[l]{\textbf{Cosmological}\\\textbf{parameters}}} & \textbf{Priors} & \textbf{Simulations} \\ \cmidrule{2-3}
         & (GWTC-4 \& O5 mock data) & \textbf{injected $\Lambda_{\rm cosmo}$} \\
        \midrule
        \multirow{2}{*}{$H_0~\rm [km \,s^{-1} \,Mpc^{-1}]$} & (0, 200) for GR models & \multirow{2}{*}{67.8} \\
         & (0, 120) for beyond GR models &  \\
        $\Omega_m$ & (0, 1) & 0.308 \\
        $w_0$ & (-3, 1) & -1 \\
        $w_a$ & (-3, 2) & 0 \\
        $\Xi_0$ & (0.435, 10) & 1 \\
        $n$ & (0.1, 10) & - \\
        $c_M$ & (-10, 50) & 0 \\
        \bottomrule
    \end{tabular}
    \caption[Priors - cosmo]{Prior probability distributions assumed on parameters of the cosmological models, for both the analysis of GWTC-4 data and O5 simulated data. All priors are uniform in the given ranges. The last column shows values of the injected parameters in the simulations at O5 sensitivity.}
    \label{tab:priors_cosmo}
\end{table}

\section{Mock data production}
\label{app:mock_data_production}

\subsection{Individual events parameter estimation}
\label{app:mock_data_production:pe}

For each of the \textit{detected} event in our simulated datasets, we estimate their parameters using \bilby~\cite{Ashton:2018jfp, Romero-Shaw:2020owr, Smith:2019ucc}. We simulate colored Gaussian noise in the detectors and consequently employ a Gaussian likelihood in frequency domain, bandpassed between $[20, 1024]~{\rm Hz}$. We use the \texttt{IMRPhenomXHM} waveform model~\cite{Garcia-Quiros:2020qpx} which includes higher angular harmonics but neglects precession (thus assuming aligned spins binaries). We generate the signal and noise at sampling frequency $2048~{\rm Hz}$, starting from $20~{\rm Hz}$. The signal duration is $t_{\rm seglen} = \Delta t_{\rm mrg} + 4~{\rm s}$, with $\Delta t_{\rm mrg}$ being the time difference between the moment the binary's frequency enters the detector's band ($20~{\rm Hz}$) and the frequency at merger. The analysis start time is set to be at $t_{\rm seglen}+1~{\rm s}$ before the geocent time of the event (see below).

We sample over the following 11 parameters: $m_1, m_2$ (primary and secondary masses, such that $m_1 > m_2$), $d_L$ (luminosity distance), $\theta_{JN}$ (inclination angle), $\psi$ (polarization angle), $\phi$ (phase of the signal), $\alpha, \delta$ (right ascension and declination, i.e.~the sky localization of the binary), $\chi_1, \chi_2$ (projected component of the spin vectors of each BH along the orbital angular momentum) and $t_{\rm geocent}$ (the time at which the merger signal reaches the Earth's center).

We employ uniform priors for the component masses $m_1$ and $m_2$, within a range mildly tailored to the injected value $m_{1, \rm inj}$, i.e. between $1~\rm M_\odot$ and an upper bound linearly interpolated from the following reference values: $60~\rm M_\odot\, (300~\rm M_\odot)$ for $m_{1, \rm inj} = 10~\rm M_\odot\, (90~\rm M_\odot)$. We employ a uniform prior for the luminosity distance $d_L$ in a range mildly tailored to the injected value $d_{L, \rm inj}$ as well, i.e. in the range $[0, 10 + 2.5 d_{L, \rm inj}]~\rm Gpc$. We assume a spin prior uniform in magnitude and orientation (see Eq.~A7 of \cite{Lange:2018pyp}), similar to what is done for real events \cite{LIGOScientific:2018mvr, LIGOScientific:2021usb, KAGRA:2021vkt, LIGOScientific:2025slb}. For the sky localization we use a prior uniform on the celestial sphere. The prior on $\cos \theta_{JN}$ is uniform in $[-1, 1]$ so that the prior on the binary orientation is uniform on the unit sphere. The priors for $\phi, \psi$ is uniform in $[0, 2\pi]$. Finally, the prior for $t_{\rm geocent}$ is uniform in the range $[t_{\rm g, inj} - 0.1~{\rm s}, t_{\rm g, inj} + 0.1~{\rm s}]$ where $t_{\rm g, inj}$ is the injected geocentric time value. 

We sample the likelihood with the \nessai sampler~\cite{nessai, Williams:2021qyt, Williams:2023ppp}. To ensure a consistent sampling of the entire parameter space, we use 2000 live points for each event. As for the hierarchical inference, we use $\texttt{dlogZ} < 0.1$ as the stopping criterion.

\subsection{Estimating selection effects}
\label{app:mock_data_production:inj}

\begin{figure}[!t]
    \centering
    \includegraphics[width=0.7\linewidth]{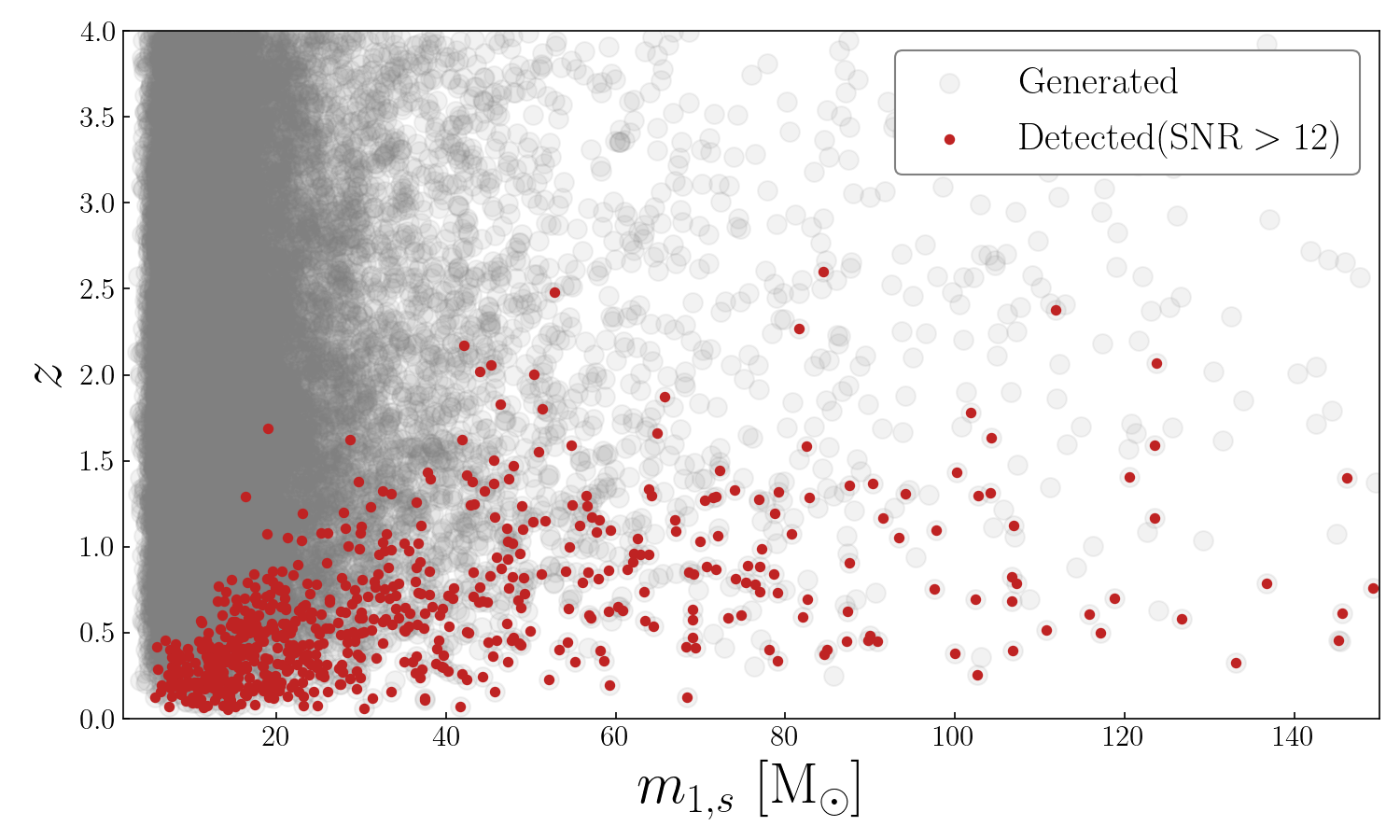}
    \caption{Subset of simulated (gray) injections at O5 sensitivity in the source-frame primary mass / redshift space. The red dots indicate those above the $\rm SNR = 12$ detection threshold.}
    \label{fig:injections}
\end{figure}

Population analyses with GWs are heavily influenced by \textit{selection effects} due to the finite detectors sensitivity~\cite{Gaebel:2018poe, Mandel:2020cig, Kapadia:2019uut, Vitale:2020aaz}. 
This effect is directly taken into account by the denominator of likelihood~\eqref{eq:hierarchical_likelihood}, which is evaluated with Monte-Carlo integration using a large set of simulated events --- normally called \textit{injections} --- mapping the events' detectability over the parameter space~\cite{Essick:2023upv}.
For \ac{GWTC} analyses, we use the publicly available injections by \ac{LVK} computed at O3 and O4 sensitivity in real noise~\cite{GWTC-4-0:injections}, that have been used in the LVK publications for GWTC-4.0~\cite{LIGOScientific:2025pvj, LIGOScientific:2025jau}. 
For simulations, we generate our own injection sets at O5 sensitivity, following a procedure similar to~\cite{Agarwal:2024hld, Bertheas:2025mzd}.
We draw events from a wide population distribution that largely covers the target one, to ensure that a sufficient number of points is available to evaluate selection effects by Monte-Carlo integration during the inference (e.g. see~\cite{Mastrogiovanni:2023zbw}).
We adopt a decreasing truncated power-law with index $\alpha = 3$ for the primary mass, between $2~{\rm M_\odot}$ and $150~{\rm M_\odot}$, smoothed at its lower bound with a wide tempering of $\delta_m = 17~{\rm M_\odot}$; a truncated power-law with index $\alpha_q = 1$ for the mass ratio; a power-law with index $\gamma = 0$ for the rate evolution.
We generate injections up to $z = 4$, noticing that, at O5 sensitivity, high-mass events can be detected up to $z \sim 3.5$.
We process the simulated injections through the same detection procedure as the mock events, computing the matched-filter \ac{SNR} in colored Gaussian noise and using $\rm SNR = 12$ as selection threshold, as described in Sec.~\ref{sec:methods:data:mock}.
Following the aforementioned method, we generate $\sim 1.7 \times 10^9$ sources and end up with $\sim 5 \cdot 10^7$ injections above the SNR threshold. 
Fig.~\ref{fig:injections} shows a small subset of the generated sources, with $\sim 350$ above detection threshold, and illustrate the positive mass -- SNR correlation. 
We comment on the numerical stability in App.~\ref{app:numerical_stability}.

\section{Stability to detection threshold}
\label{app:far_comparison}

\begin{figure}[!b]
    \centering
    \begin{subfigure}[b]{0.49\textwidth}
        \centering
        \includegraphics[width=\linewidth]{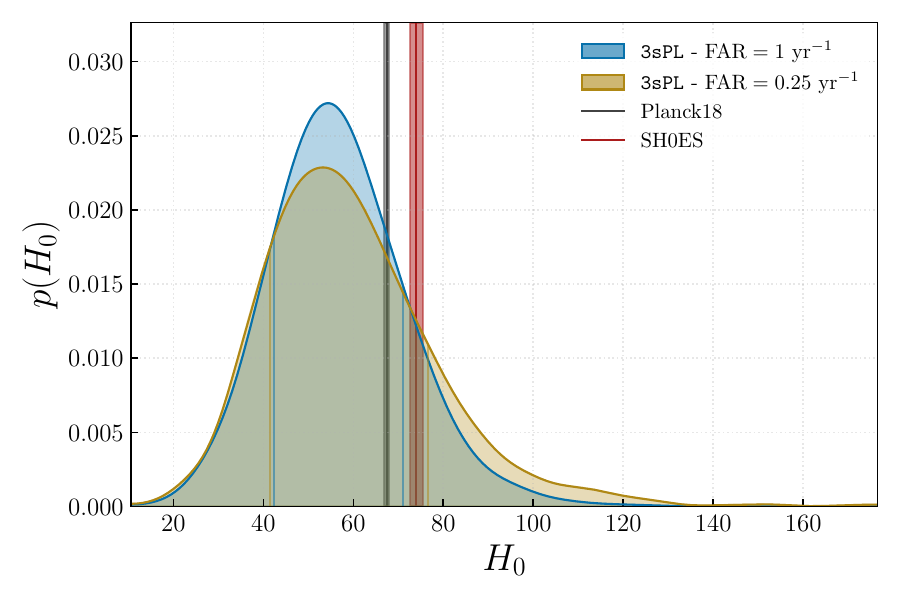}
        \caption{}
        \label{fig:far_comparison_H0}
    \end{subfigure}
    \begin{subfigure}[b]{0.49\textwidth}
        \centering
        \includegraphics[width=\linewidth]{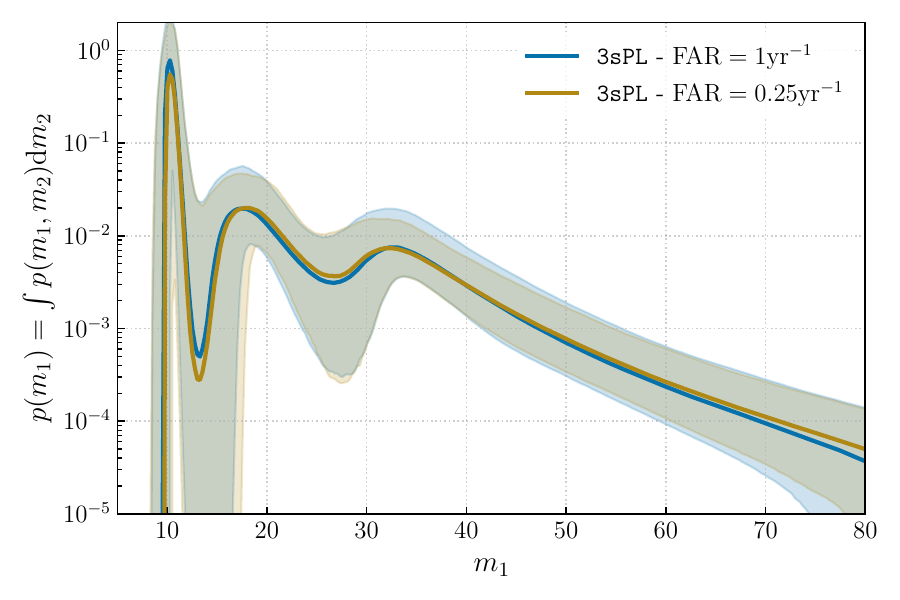}
        \caption{}
        \label{fig:far_comparison_PPD}
    \end{subfigure}
    \caption[GWTC-4.0 - FAR comparison]{Comparison of \camelthrees $H_0$ constraints \textbf{(a)} and primary mass posterior predictive distribution \textbf{(b)} from our $\rm FAR < 1~yr^{-1}$ dataset (blue) and LVK's $\rm FAR < 0.25~yr^{-1}$ dataset.}
    \label{fig:far_comparison}
\end{figure}
In this section, we present a brief comparison of analyses results in terms of $H_0$ constraints and \acp{PPD} in a \LCDM scenario. Fig.~\ref{fig:far_comparison} compares a run with the GWTC-4.0 dataset used throughout the present paper, composed of BBH events with $\rm FAR < 1~yr^{-1}$ excluding two high mass and one mass-asymmetric events as described in Sec.~\ref{sec:methods}, with a run performed with the same event list as in the \ac{LVK} cosmology paper \cite{LIGOScientific:2025jau}, which adopts a cut of $\rm FAR < 0.25~yr^{-1}$, including GW231123, GW190521 and GW190412. We note no major difference between the two: \acp{PPD} of primary mass in Fig.~\ref{fig:far_comparison_PPD} are identical, and $H_0$ posteriors in Fig.~\ref{fig:far_comparison_H0} are perfectly compatible, proving that our parametrisation is robust against the selection criterion. We even note a (very) small improvement of $H_0$ constraints when using our $\rm FAR < 1~yr^{-1}$ dataset, which contains 13 events more than \ac{LVK}'s.

\section{Numerical stability}
\label{app:numerical_stability}

The \icarogw implementation of the hierarchical likelihood evaluates its various integrals with \ac{MC} techniques~\cite{Mastrogiovanni:2023zbw}, using both posterior samples from individual events and injections to compute selection effects. 
We assess numerical stability of the likelihood by evaluating the variance of the estimator of its logarithm $\sigma^2_{\ln \mathcal{\hat L}}$, and accept only samples of population parameters with $\sigma^2_{\ln \mathcal{\hat L}} < 1$, as recommended in~\cite{Talbot:2023pex}. 
This numerical stability criterion is stricter than thresholds on effective numbers of \ac{PE} samples $N_{\rm eff, PE} > 10$ and injections $N_{\rm eff, inf} > 4 \times N_{\rm obs}$ used in various related works~\cite{LIGOScientific:2021aug, LIGOScientific:2025jau, Gennari:2025nho, Bertheas:2025mzd}.
This choice ensures more stable handling of potentially sharp features in the mass distribution as those observed in Sec.~\ref{sec:res:gwtc4}.

\begin{table}[htb!]
    \centering
    \begin{tabular}{@{\extracolsep\fill}lcccccc}
        \toprule%
        \makecell[r]{\textbf{Population $\rightarrow$}} & \multicolumn{2}{c}{\multirow{2}{*}{\camelthrees}} & \multicolumn{2}{c}{\multirow{2}{*}{\camelfours}} & \multicolumn{2}{c}{\multirow{2}{*}{\sPLGG}} \\ 
        \textbf{$\downarrow$ Cosmology} & & & & & & \\
        \midrule
        \texttt{FlatLCDM} & $0$ & $(0)$ & $-0.6$ & $(-1.1)$ & $-0.8$ & $(+0.5)$ \\
        \texttt{FlatLCDM} (fix $\Omega_m$) & $+0.2$ & $(+0.2)$ & $-0.6$ & $(-1.1)$ & $-0.7$ & $(+0.6)$ \\ \cmidrule{1-1}
        \texttt{w0waCDM} & $-0.1$ & $(-0.2)$ & $-0.6$ & $(-1)$ & $-0.8$ & $(+0.5)$ \\ \cmidrule{1-1}
        \texttt{Xi0} (fix $\Omega_m$) & $-0.2$ & $(-0.3)$ & $-1$ & $(-1.4)$ & $-1.3$ & $(-0.1)$ \\
        \texttt{Xi0} (fix $H_0, \Omega_m$) & $0$ & $(-0.2)$ & $-1$ & $(-1.3)$ & $-1.1$ & $(+0.1)$ \\ \cmidrule{1-1}
        \texttt{cM} (fix $\Omega_m$) & $-0.2$ & $(-0.7)$ & $-1.1$ & $(-1.8)$ & $-1.3$ & $(-0.3)$ \\
        \texttt{cM} (fix $H_0, \Omega_m$) & $-0.4$ & $(-0.8)$ & $-1$ & $(-1.5)$ & $-1.4$ & $(-0.5)$ \\
        \bottomrule
    \end{tabular}
    \caption[GWTC-4 - \acp{BF}]{\acp{BF} $\mathcal{B}$ comparing analyses results on GWTC-4 data. The values are always given \ac{BF} with respect to the \camelthrees + \texttt{FlatLCDM} analysis. For each population model, the left \ac{BF} column reports the bare \acp{BF} output by the nested sampler, while the right column in parentheses reports the \acp{BF} corrected by the effective prior volume following the prescription in section III.A of~\cite{Mould:2025dts}.}
    \label{tab:gwtc4_BFs}
\end{table}

Nested samplers estimate the evidence from the region of the parameter space that they explore during the inference process. In our case, however, part of this region is excluded from the analysis due to numerical stability constraints. Consequently, the resulting evidence corresponds to the integral of a smaller volume of the full problem. To compensate for this effect, previous studies have suggested reweighting the evidence using the effective stable volume~\cite{Mould:2025dts}. We report the corrected Bayes factors in Tab.~\ref{tab:gwtc4_BFs} following this procedure. However, we argue that this approximation is not helpful in our case, where the excluded regions coincide with those of high likelihood~\cite{Gennari:2025nho}. In such a scenario, excluding an informative region from the analysis cannot simply be compensated for by volume effects, and developing new techniques will be required to solve this problem robustly.
We nonetheless observe that the correction does not have an important effect on relative evidences, although \camelthrees is not particularly favored over \sPLGG anymore: the two models exhibit almost equal evidence instead. 
Notice also that according to this corrected metric, \camelfours is still mildly disfavoured with respect to \camelthrees and \sPLGG.



\bibliographystyle{JHEP}
\bibliography{references}

\end{document}